\documentclass[conference]{IEEEtran}
\IEEEoverridecommandlockouts
\usepackage{cite}
\usepackage{amsmath,amssymb,amsfonts}
\usepackage{algorithmic}
\usepackage{graphicx}
\usepackage{textcomp}
\usepackage{xcolor}
\usepackage{booktabs}
\usepackage{multirow}
\usepackage{booktabs}
\usepackage[table]{xcolor}
\usepackage{amsfonts}
\usepackage{hyperref}
\usepackage{cleveref}

\definecolor{lightgray}{RGB}{240,240,240}
\definecolor{lightblue}{RGB}{240,248,255}
\definecolor{lightyellow}{RGB}{255,250,205}

\definecolor{lightgreen}{HTML}{EAF2E3}
\definecolor{lightorange}{HTML}{FFF3E6}
\def\BibTeX{{\rm B\kern-.05em{\sc i\kern-.025em b}\kern-.08em
    T\kern-.1667em\lower.7ex\hbox{E}\kern-.125emX}}
\begin{document}

\title{
% Uncertainty Estimation in Audio-Aware Large Language Models: A Systematic Empirical Study
Walking Through Uncertainty: \\
An Empirical Study of Uncertainty Estimation for Audio-Aware Large Language Models
}

% Walking Through Uncertainty: An Empirical Study of Uncertainty Estimation for Audio-Aware Large Language Models

% Walking Through Uncertainty: An Empirical Study of Uncertainty Estimation in Audio-Aware Large Language Models

% Walking Through Uncertainty: A Systematic Empirical Study of Uncertainty Estimation in Audio-Aware Large Language Models
% Walking Through Uncertainty: An Empirical Study of Uncertainty Estimation in Audio-Aware Large Language Models
% Walking Through Uncertainty: A Systematic Study of Uncertainty Estimation in Audio-Aware Large Language Models
% Walking Through Uncertainty in Audio-Aware Large Language Models: A Systematic Empirical Study

% \title{Uncertainty Estimation in Audio-Aware Large Language Models: An Empirical Study}

% Uncertainty Estimation in Audio-Aware Large Language Models: A Systematic Evaluation Across Understanding and Trustworthiness Tasks

% An Empirical Study of Uncertainty Estimation in Audio-Aware Large Language Models

% \author{
% Chun-Yi Kuan, Wei-Ping Huang, Hung-yi Lee \\
% National Taiwan University, NTU Artificial Intelligence Center of Research Excellence \\
% chunyi.kuan.tw@gmail.com
% }

% \author{
% Chun-Yi Kuan$^{\heartsuit}$, 
% Wei-Ping Huang$^{\heartsuit}$, 
% Hung-yi Lee$^{\heartsuit}$$^{\clubsuit}$ \\
% $^{\heartsuit}$National Taiwan University \\
% $^{\clubsuit}$NTU Artificial Intelligence Center of Research Excellence (NTU AI-CoRE) \\
% % $^{\clubsuit}$Artificial Intelligence Center of Research Excellence, National Taiwan University \\
% chunyi.kuan.tw@gmail.com
% }

\author{
Chun-Yi Kuan$^{\heartsuit}$, 
Wei-Ping Huang$^{\heartsuit}$, 
Hung-yi Lee$^{\heartsuit}$$^{\clubsuit}$ \\
$^{\heartsuit}$Graduate Institute of Communication Engineering, National Taiwan University, Taiwan  \\
$^{\clubsuit}$Artificial Intelligence Center of Research Excellence (AI-CoRE), National Taiwan University, Taiwan \\
% $^{\clubsuit}$Artificial Intelligence Center of Research Excellence, National Taiwan University \\
chunyi.kuan.tw@gmail.com
}

\maketitle

\begin{abstract}
Recent audio-aware large language models (ALLMs) have demonstrated strong capabilities across diverse audio understanding and reasoning tasks, but they still frequently produce hallucinated or overly confident outputs. 
While uncertainty estimation has been extensively studied in text-only LLMs, it remains largely unexplored for ALLMs, where audio-conditioned generation introduces additional challenges such as perceptual ambiguity and cross-modal grounding. 
In this work, we present the first systematic empirical study of uncertainty estimation in ALLMs. 
We benchmark five representative methods, including predictive entropy, length-normalized entropy, semantic entropy, discrete semantic entropy, and P(True), across multiple models and diverse evaluation settings spanning general audio understanding, reasoning, hallucination detection, and unanswerable question answering. 
Our results reveal two key findings. 
First, semantic-level and verification-based methods consistently outperform token-level baselines on general audio reasoning benchmarks. 
Second, on trustworthiness-oriented benchmarks, the relative effectiveness of uncertainty methods becomes notably more model- and benchmark-dependent, indicating that conclusions drawn from general reasoning settings do not straightforwardly transfer to hallucination and unanswerable-question scenarios. 
We further explore uncertainty-based adaptive inference as a potential downstream application. 
We hope this study provides a foundation for future research on reliable, uncertainty-aware audio-language systems.
\end{abstract}


\section{Introduction}

Recent audio-aware large language models (ALLMs)~\cite{achiam2023gpt, gong2023listen, gong2023joint, wang2023blsp, fathullah2023towards, kuan2024speech, chang2024speechprompt, wang2024blsp, goel2025audio, liu2025voxtral, xu2025qwen2, yang2024building, kuan2025teaching, kuan2025alignment, guo2025recentadvancesdiscretespeech, wu2024audiolanguagemodeling, mousavi2025discreteaudiotokenssurvey, cui2025recentadvancesspeechlanguage, peng2025surveyspeechlargelanguage, ji2024wavchatsurveyspokendialogue, abouelenin2025phi, comanici2025gemini, arora2025landscape, ghoshaudio, chang2026tico} have rapidly advanced across a wide range of tasks, including audio understanding~\cite{sakshimmau, ma2025mmar, kumar2025mmau, wang2025mmsu, yang2025sakura, kuan2024understanding, kuan2025can}, spoken question answering~\cite{huang2024dynamic, huang2024dynamic2, wang2025mmsu}, music reasoning
, and general audio-language interaction~\cite{lin2024listen, lin2024spoken, tseng2024av, chen2024beyond, he2025audiomarathon, lu2025speech, tseng2025evaluation, yang2025paras2s, ren2026mos, kuan2026aqua, kuan2026aqascore, chen2026causal, dong2026membership, zhou2025echomind, lin2026contrastive, chang2026taigispeech, lin2026vibe, chang2025game}. 
By conditioning language generation on audio inputs, these models extend the capabilities of text-only LLMs to richer multimodal scenarios. 
However, strong performance does not necessarily imply reliability. 
In practice, ALLMs still frequently produce unsupported answers~\cite{chen2025aha}, hallucinated content~\cite{kuan2024understanding, kuan2025can, chen2025aha, zhao2026halluaudio, li2025audiotrust}, or overly confident responses~\cite{he2025measuring, kuan2026aqua}, especially when the audio evidence is ambiguous, incomplete, or even insufficient to answer the question.

This limitation highlights a fundamental gap: beyond accuracy, we need to understand whether a model knows when it may be wrong. 
Uncertainty estimation~\cite{hendrycks2016baseline, ovadia2019can, desai2020calibration, rae2021scaling, jiang2021can, kadavath2022language, mielke2022reducing, srivastava2023beyond, kuhnsemantic, farquhar2024semanticentropy, nikitin2024kernel, nguyen2025beyond, wangself, zhu2023calibration, guo2017calibration, lin2022teaching, yin2023large, chen2025uncertainty, huang2024survey, hu2023uncertainty, xia2025survey, aichberger2024rethinking, chen2025seed} provides a natural framework for addressing this problem. 
Reliable uncertainty estimates can support error detection, selective prediction, calibration, and safer deployment. 
In text-only LLMs, a large body of work~\cite{kadavath2022language, kuhnsemantic, farquhar2024semanticentropy, nikitin2024kernel, nguyen2025beyond, wangself, zhu2023calibration, guo2017calibration, lin2022teaching, yin2023large, chen2025uncertainty, huang2024survey, hu2023uncertainty, xia2025survey, aichberger2024rethinking, chen2025seed} has explored uncertainty estimation through predictive entropy, semantic entropy~\cite{kuhnsemantic}, and self-verification-based measures such as P(True)~\cite{kadavath2022language}, showing that uncertainty can serve as an effective signal for both analysis and downstream decision-making.

However, it remains unclear whether these insights transfer to audio-aware LLMs. 
Although ALLMs inherit language modeling capabilities from their text backbones, audio-conditioned generation introduces additional challenges. 
The model must reason over noisy and imperfect perceptual inputs, ground its predictions in acoustic evidence, and avoid over-reliance on language priors when the audio signal is weak or ambiguous. 
As a result, uncertainty signals that are effective in text-only settings might not behave the same when correctness depends on both perception and reasoning.

In this work, we present the first systematic empirical study of uncertainty estimation in audio-aware LLMs. 
We evaluate several representative methods, including predictive entropy, length-normalized entropy, semantic entropy~\cite{kuhnsemantic}, discrete semantic entropy~\cite{kuhnsemantic}, and P(True)~\cite{kadavath2022language}, across multiple state-of-the-art ALLMs~\cite{xu2025qwen2, goel2025audio} and a diverse set of benchmarks~\cite{sakshimmau, ma2025mmar, wang2025mmsu, yang2025sakura, kuan2024understanding, kuan2026aqua}.
Our evaluation spans both general audio understanding and reasoning tasks~\cite{sakshimmau, ma2025mmar, wang2025mmsu, yang2025sakura}, and more challenging trustworthiness-oriented settings, including hallucination detection~\cite{kuan2024understanding, kuan2025can} and unanswerable audio question answering~\cite{kuan2026aqua}.
Our results reveal two key findings. 
First, on general audio understanding and reasoning benchmarks, semantic-level and verification-based methods consistently outperform token-level likelihood-based baselines. 
This suggests that uncertainty signals that operate at the level of meaning or explicit self-evaluation are better aligned with correctness in audio-language tasks. 
Second, this advantage becomes notably less stable on trustworthiness-oriented benchmarks, where the relative performance of uncertainty estimation methods is more strongly dependent on the model and the task. 
This indicates that conclusions drawn from standard reasoning benchmarks do not directly transfer to settings that explicitly test reliability, such as hallucination and unanswerable questions.
Finally, we explore uncertainty-based adaptive inference~\cite{chen2023frugalgpt, aggarwal2023automix, zhang2025leveraging, moslem2026dynamic} as a possible downstream application.
We show that uncertainty can be used as a routing signal to selectively trigger more expensive reasoning strategies. 
However, its effectiveness critically depends on whether the alternative reasoning mode is itself beneficial~\cite{ma2025audio, xie2025audio, fan2025incentivizing, patel2025sound, zhang2025adaptthink, snell2024scaling, bilal2026if, ghosal2025does, qu2026adaptive, li2025think}. 
This highlights that uncertainty alone is not sufficient, as its utility is tightly coupled with the underlying inference strategy.

Overall, this work provides the first comprehensive evaluation of uncertainty estimation in audio-aware LLMs, establishes empirical baselines, and reveals key differences between general reasoning and trustworthiness-oriented settings. 
We hope these findings provide a foundation for developing more reliable and uncertainty-aware audio-language systems.

\section{Related Work}

\subsection{Uncertainty Estimation in Language Models}

% Prior work has studied uncertainty estimation in language models as a way to improve the reliability of generated outputs. 
% Existing approaches can be broadly categorized into three groups. 
% First, token-level methods estimate uncertainty from the predictive distribution over generated sequences, such as predictive entropy and normalized variants. 
% Second, semantic-level methods argue that uncertainty in natural language generation should be measured modulo semantic equivalence, since multiple responses may differ lexically while expressing the same meaning. 
% A representative example is semantic entropy, which clusters semantically equivalent generations before computing entropy. 
% Third, verification-based methods estimate confidence by asking the model to assess whether its own answer is correct, as in P(True).
% Beyond measuring confidence itself, prior work has shown that uncertainty estimates can support downstream decision-making, including calibration, hallucination detection, and selective prediction. 
% Despite this growing literature in text-based language models, uncertainty estimation remains much less studied in audio-aware LLMs. 
% This leaves open the question of whether methods developed for text-only generation transfer reliably to audio-conditioned settings.

Uncertainty estimation has become an important topic in language modeling, as large language models (LLMs) can produce fluent yet unreliable outputs. 
Prior work~\cite{kadavath2022language, kuhnsemantic, farquhar2024semanticentropy, nikitin2024kernel, nguyen2025beyond, wangself, zhu2023calibration, guo2017calibration, lin2022teaching, yin2023large, chen2025uncertainty, huang2024survey, hu2023uncertainty, xia2025survey, aichberger2024rethinking, chen2025seed} has explored a range of uncertainty estimation methods, which can be broadly grouped into token-level, semantic-level, and verification-based approaches. 
Token-level methods quantify uncertainty from the predictive distribution over generated sequences, for example through predictive entropy and related likelihood-based measures. 
Semantic-level methods~\cite{kuhnsemantic} instead account for the fact that multiple generations may differ in surface form while conveying the same meaning, and therefore measure uncertainty over clusters of semantically equivalent responses. 
Verification-based approaches~\cite{kadavath2022language}, such as P(True), estimate confidence by prompting the model to assess whether its own answer is correct.
Beyond measuring confidence itself, prior work has shown that uncertainty estimates can support downstream decision-making, including calibration~\cite{guo2017calibration, kadavath2022language, geng2024survey}, hallucination detection~\cite{farquhar2024semanticentropy, manakul2023selfcheckgpt}, and selective prediction~\cite{varshney-etal-2022-investigating, feng2024don, wen2025know}. 
In particular, semantic entropy has been shown to be effective for detecting hallucinations~\cite{farquhar2024semanticentropy} in natural language generation by measuring uncertainty modulo semantic equivalence. 
Similarly, self-evaluation-based confidence estimates such as P(True) have been used to assess whether a model's generated answer is likely to be correct, and subsequent work has shown that self-evaluation can further improve selective generation in LLMs.
More broadly, uncertainty estimates have also been used in downstream settings such as hallucination detection, selective prediction, and adaptive generation.

Despite this growing literature, most existing findings are established in text-only language models. 
Whether these uncertainty estimation methods remain effective in audio-aware LLMs is still unclear, since audio-conditioned generation introduces additional challenges such as perceptual ambiguity, cross-modal grounding, and reliance on language priors. 
Moreover, the use of uncertainty estimates for downstream decision-making remains much less explored in audio-aware LLMs. 
Our work builds on this line of research and systematically investigates how uncertainty estimation methods behave in audio-aware LLMs.

\subsection{Audio-aware LLM Evaluation and Trustworthiness}

As audio-aware LLMs have rapidly advanced, recent work has introduced increasingly diverse benchmarks to evaluate their capabilities. 
Existing benchmarks cover general audio understanding and reasoning~\cite{sakshimmau, huang2024dynamic, huang2024dynamic2, he2025audiomarathon, kumar2025mmau, lu2025speech, ma2025mmar, tseng2025evaluation, wang2025mmsu, yang2025paras2s, yang2025sakura, zhou2025echomind, chang2026taigispeech, chen2026causal, dong2026membership, kuan2026aqascore, lin2026contrastive, lin2026vibe, ren2026mos} across speech, environmental sounds, and music, with recent efforts expanding evaluation toward more challenging reasoning settings such as expert-level audio understanding and multi-hop reasoning~\cite{kumar2025mmau, ma2025mmar, yang2025sakura, zhou2025echomind}. 
These developments have substantially improved the empirical coverage of audio-aware LLM evaluation.
At the same time, recent work has begun to highlight trustworthiness issues~\cite{kuan2024understanding, lin2024listen, lu2025speech, kuan2025can, lin2026vibe, chen2025aha, li2025audiotrust, zhao2026halluaudio, kuan2026aqua} in audio-aware LLMs. 
Prior studies~\cite{kuan2024understanding, kuan2025can} show that these models can hallucinate sound events that are not present in the input audio, revealing a gap between surface-level fluency and faithful audio grounding~\cite{kuan2024understanding}. 
More recent benchmarks~\cite{kuan2026aqua} extend this line of work by explicitly evaluating unanswerable audio questions and other trustworthiness-oriented settings. 
Together, these works suggest that evaluating audio-aware LLMs should not be limited to answer accuracy alone, but should also consider whether model outputs are reliable, grounded, and appropriately calibrated with respect to the input audio.
Despite this growing interest in evaluating audio-aware LLM capabilities and trustworthiness, uncertainty estimation remains largely underexplored in this setting. 
Existing studies primarily assess what models get right or wrong, but provide limited insight into whether models can reliably indicate when they may be wrong. 
Our work addresses this gap by systematically studying uncertainty estimation in audio-aware LLMs across both general reasoning and trustworthiness-oriented benchmarks.
Beyond benchmark evaluation, it also remains unclear whether uncertainty estimates can support trustworthiness-oriented decision making in audio-aware LLMs, such as hallucination detection or adaptive inference.

% \subsection{Uncertainty-aware Decision Making}

\section{Method}

\subsection{Problem Setup}

Given an input $x$ consisting of an audio clip and a corresponding instruction or question, an audio-aware LLM generates an answer $y$ according to the conditional distribution $p(y \mid x)$.
Our goal is to estimate how uncertain the model is about its prediction. 
Therefore, we compare several representative uncertainty estimation methods, including predictive entropy, length-normalized predictive entropy, semantic entropy~\cite{kuhnsemantic}, discrete semantic entropy~\cite{kuhnsemantic}, and P(True)~\cite{kadavath2022language}.
% These methods differ in the level at which uncertainty is measured. 
Predictive entropy and its length-normalized variant operate on the likelihood of generated sequences, semantic entropy measures uncertainty over semantically equivalent answer clusters, discrete semantic entropy estimates uncertainty over a discrete answer space, and P(True) measures the model's self-assessed probability that its answer is correct.

\subsection{Prediction and Uncertainty Estimation Protocol}
\label{method/protocol}
Following common practice~\cite{kadavath2022language, kuhnsemantic} in sampling-based uncertainty estimation for language models, we adopt a two-stage protocol that decouples prediction from uncertainty estimation. 
Given an input $x$, we first obtain a prediction $\hat{y}$ using low-temperature decoding, which serves as the model's final answer for correctness evaluation.
We then estimate uncertainty by sampling multiple responses from the predictive distribution $p(y \mid x)$ and computing uncertainty measures over the resulting samples.
Formally, let $\hat{y}$ denote the prediction obtained via low-temperature decoding. 
We evaluate whether the model is correct based on $\hat{y}$. 
Separately, we draw $K$ stochastic samples $\{y_i\}_{i=1}^{K}$ from $p(y \mid x)$ and use them to estimate uncertainty, such as predictive entropy, length-normalized entropy, semantic entropy, and discrete semantic entropy.
For P(True), instead of estimating uncertainty from sampled responses, we directly verify the low-temperature prediction $\hat{y}$ by prompting the model to judge whether $\hat{y}$ is supported by the input $x$.

\subsection{Uncertainty Estimation Methods}

\textbf{Predictive entropy} measures the overall uncertainty of the model's generated answer under the conditional distribution $p(y \mid x)$. Formally, the sequence-level predictive entropy is defined as
\begin{equation}
H_{\text{pred}}(x) = \mathbb{E}_{y \sim p(y \mid x)} \left[ - \log p(y \mid x) \right].
\end{equation}
Since the exact expectation is intractable, we approximate it with Monte Carlo sampling. Given $K$ sampled answers $\{y_i\}_{i=1}^{K}$ from $p(y \mid x)$, we estimate
\begin{equation}
H_{\text{pred}}(x)
\approx
\frac{1}{K} \sum_{i=1}^{K} - \log p(y_i \mid x).
\end{equation}
A higher predictive entropy indicates that the model assigns lower probability to its sampled generations on average, reflecting greater uncertainty at the sequence level.
Predictive entropy can be biased by answer length, since longer sequences typically accumulate larger negative log-likelihoods. 
To reduce this effect, we additionally consider a \textbf{length-normalized predictive entropy} that measures average uncertainty per token:
\begin{equation}
H_{\text{norm}}(x)
=
\mathbb{E}_{y \sim p(y \mid x)}
\left[
\frac{- \log p(y \mid x)}{|y|}
\right],
\end{equation}
where $|y|$ denotes the number of tokens in $y$.
A straightforward Monte Carlo estimator is
\begin{equation}
H_{\text{norm}}(x)
\approx
\frac{1}{K}
\sum_{i=1}^{K}
\frac{- \log p(y_i \mid x)}{|y_i|}.
\end{equation}

In practice, however, we adopt a token-level estimator by normalizing the total negative log-likelihood with the total number of generated tokens:
\begin{equation}
H_{\text{norm}}^{\text{tok}}(x)
=
\frac{
\sum_{i=1}^{K} - \log p(y_i \mid x)
}{
\sum_{i=1}^{K} |y_i|
}.
\end{equation}
This formulation can be viewed as a ratio-of-expectations approximation, and effectively estimates the average token-level uncertainty across sampled responses. Compared with per-sequence normalization, it is more stable in practice and less sensitive to variance induced by different answer lengths.

Sequence-level likelihood does not distinguish between surface-form variation and genuine semantic uncertainty. Multiple sampled answers may differ lexically while expressing the same meaning. To capture uncertainty at the semantic level, we follow and cluster sampled answers by semantic equivalence.
Specifically, semantic equivalence is determined via bidirectional entailment clustering, following the original implementation.
Given two sampled answers $s$ and $s'$, we concatenate each answer with the original context $x$ and use a natural language inference (NLI) model to test whether $(x, s)$ entails $(x, s')$ and vice versa. Two answers are assigned to the same semantic cluster if and only if both entailment relations hold.
Let $\{c_k\}$ denote the resulting semantic clusters. The probability mass of a cluster is obtained by summing the sequence probabilities of all answers assigned to that cluster:
\begin{equation}
p(c_k \mid x) = \sum_{s \in c_k} p(s \mid x).
\end{equation}
The \textbf{semantic entropy} is then defined as
\begin{equation}
H_{\text{sem}}(x)
=
- \sum_k p(c_k \mid x)\log p(c_k \mid x).
\end{equation}
Intuitively, semantic entropy is low when most sampled answers collapse into the same meaning cluster, and high when the model spreads probability mass across multiple semantically distinct answers.
% For tasks with a discrete answer space, such as multiple-choice question answering, uncertainty can be estimated directly over the model's empirical answer distribution. 
% Let $\mathcal{A}$ denote the set of candidate answer options. 
% Given $K$ sampled answers, we estimate the probability of choosing option $a \in \mathcal{A}$ as
% \begin{equation}
% p(a \mid x)
% \approx
% \frac{\mathrm{count}(a)}{K}.
% \end{equation}
% The \textbf{discrete semantic entropy} is then computed as
% \begin{equation}
% H_{\text{disc}}(x)
% =
% - \sum_{a \in \mathcal{A}} p(a \mid x) \log p(a \mid x).
% \end{equation}
While semantic entropy measures uncertainty over clusters of semantically equivalent free-form answers, some tasks involve a predefined discrete answer space. 
% Discrete semantic entropy can be viewed as a simplified variant of semantic entropy for tasks with a closed-form answer space.
In such cases, semantic variation is naturally constrained by the candidate answer set, and uncertainty can be estimated directly over the model's empirical distribution on answer choices. We refer to this variant as \textbf{discrete semantic entropy}.
Let $\mathcal{A}$ denote the set of candidate answers. Given $K$ sampled responses for the same input $x$, we map each response to an answer option in $\mathcal{A}$ and estimate the empirical answer distribution as
\begin{equation}
p(a \mid x)
\approx
\frac{\mathrm{count}(a)}{K},
\quad a \in \mathcal{A}.
\end{equation}

The discrete semantic entropy is then defined as
\begin{equation}
H_{\text{disc}}(x)
=
- \sum_{a \in \mathcal{A}} p(a \mid x)\log p(a \mid x).
\end{equation}
This quantity is low when the sampled answers concentrate on a single option, and high when the model spreads probability mass across multiple candidate answers. 
It can be viewed as a discrete counterpart of semantic entropy for closed-form answer spaces, avoiding the need for semantic equivalence clustering.

In addition to entropy-based methods, we consider \textbf{P(True)}, a self-verification-based confidence estimate. 
Given an input $x$ and a candidate answer $\hat{y}$, the model is prompted to judge whether the answer is supported by the input. 
In our setting, this corresponds to self-verification conditioned on the audio input.
Formally, we define
\begin{equation}
P(\mathrm{true} \mid x, \hat{y})
=
p(\text{true} \mid x, \hat{y})
\end{equation}
where the model predicts whether $\hat{y}$ is correct or supported by the input.
In practice, we instantiate P(True) with a self-verification prompt that asks the model to judge whether a candidate answer is supported by the input audio. 
The model is constrained to produce a binary decision (\texttt{true} or \texttt{false}), and the confidence score is computed from the normalized likelihood of these two verification tokens. 
The full prompt templates used in our experiments are provided in Table~\ref{tab:ptrue_prompt}.
A higher P(True) indicates that the model assigns higher confidence to the correctness of its own answer. 
In this work, we use self-verification only, without relying on an external verifier model.

Overall, these methods quantify uncertainty at different levels: predictive entropy and length-normalized entropy measure uncertainty from sequence likelihoods, semantic entropy and discrete semantic entropy capture dispersion over meaning or answer choices, and P(True) estimates confidence through explicit self-verification.

% In practice, we compute this probability by normalizing the likelihoods of the binary verification tokens:
% \begin{equation}
% P(\text{true} \mid x, \hat{y})
% =
% \frac{
% \exp \left( \log p(\text{``true''} \mid x, \hat{y}) \right)
% }{
% \exp \left( \log p(\text{``true''} \mid x, \hat{y}) \right)
% +
% \exp \left( \log p(\text{``false''} \mid x, \hat{y}) \right)
% }.
% \end{equation}

% Equivalently, this can be written as a sigmoid over the log-likelihood ratio:
% \begin{equation}
% P(\text{true} \mid x, \hat{y})
% =
% \sigma \left(
% \log p(\text{``true''} \mid x, \hat{y})
% -
% \log p(\text{``false''} \mid x, \hat{y})
% \right).
% \end{equation}

\section{Experimental Setup}

\subsection{Evaluation Datasets, Models, and Setups}

We evaluate uncertainty estimation across diverse benchmarks and representative audio-aware LLMs. 
Our evaluation covers two complementary settings. 
The first consists of general audio understanding and reasoning benchmarks, including MMAU~\cite{sakshimmau}, MMAR~\cite{ma2025mmar}, MMSU~\cite{wang2025mmsu}, and SAKURA~\cite{yang2025sakura}. 
The second focuses on trustworthiness-oriented evaluation, including audio hallucination and unanswerable question answering benchmarks such as Audio-Hallucination~\cite{kuan2024understanding} and AQUA-Bench~\cite{kuan2026aqua}.
We follow the evaluation protocols defined in each benchmark to ensure consistency with prior work.
We conduct experiments on several representative audio-aware LLMs, including Qwen2.5-Omni-3B~\cite{xu2025qwen2}, Qwen2.5-Omni-7B~\cite{xu2025qwen2}, and Audio Flamingo 3~\cite{goel2025audio}. 
These models are chosen for their strong performance and broad use in recent audio-language studies.

For the two-stage protocol described in Section~\ref{method/protocol}, we follow common practice in sampling-based uncertainty estimation for language models. 
Specifically, we use low-temperature decoding with temperature $0.1$ to obtain the prediction $\hat{y}$ for correctness evaluation, so that $\hat{y}$ more closely reflects the model's most likely answer. 
For sampling-based uncertainty estimation, we draw $K=10$ stochastic samples using temperature $1.0$ to better characterize the predictive distribution.
The evaluation prompt template is shown in Table~\ref{tab:basic_inference_prompt}.
In all experiments, inference is performed using a single NVIDIA RTX 3090 GPU.

\subsection{Evaluation Metrics}

We evaluate uncertainty estimation using two metrics: \textit{AUROC} and \textit{AURAC}.
We use the area under the receiver operating characteristic curve (AUROC) to measure how well an uncertainty estimate separates correct predictions from incorrect ones. 
In our setting, each example is associated with a correctness label and an uncertainty score. 
A good uncertainty estimate should assign higher uncertainty to incorrect predictions and lower uncertainty to correct ones. 
AUROC summarizes this ranking quality across all possible decision thresholds. 
A higher AUROC indicates that the uncertainty score is more effective at distinguishing errors from correct answers.
We also use the area under the rejection--accuracy curve (AURAC) to evaluate the usefulness of uncertainty estimates for selective prediction. 
The rejection--accuracy curve plots the model accuracy after progressively rejecting the most uncertain examples. 
If an uncertainty estimate is reliable, removing high-uncertainty predictions should lead to a faster increase in accuracy on the remaining examples. 
A higher AURAC therefore indicates that the uncertainty estimate is more useful for identifying predictions that should be abstained from or handled with additional care. 
We also report the original accuracy on each benchmark as a reference, since it reflects the model's default performance without abstention. 
Comparing original accuracy with AURAC helps clarify whether the uncertainty estimate provides meaningful practical utility for selective prediction beyond the model's base accuracy. 
Overall, AUROC captures the error-detection ability of an uncertainty estimate, whereas AURAC reflects its practical utility under selective prediction.
% A higher AURAC therefore indicates that the uncertainty estimate is more useful for identifying predictions that should be abstained from or handled with additional care.
% Overall, AUROC captures the error-detection ability of an uncertainty estimate, whereas AURAC reflects its practical utility under selective prediction.

% \subsection{Evaluation Datasets}

% We evaluate uncertainty estimation on several representative datasets spanning two complementary settings. 
% The first consists of general audio understanding and reasoning benchmarks, including MMAU~\cite{sakshimmau}, MMAR~\cite{ma2025mmar}, MMSU~\cite{wang2025mmsu}, and SAKURA~\cite{yang2025sakura}. 
% The second focuses on trustworthiness-oriented evaluation, including audio hallucination and unanswerable question answering benchmarks such as Audio-Hallucination~\cite{kuan2024understanding} and AQUA-Bench~\cite{kuan2026aqua}.

% \subsection{Evaluated Models}

% We evaluate uncertainty estimation on several representative audio-aware LLMs, including Qwen2.5-Omni-3B, Qwen2.5-Omni-7B, and Audio Flamingo 3. 
% These models are chosen for their strong performance and broad use in recent audio-language studies.

\section{Results and Analysis}
\begin{table*}[t]
\centering
\caption{
Uncertainty estimation results on \textbf{Understanding \& Reasoning} benchmarks.
The best AUROC in each row is highlighted with a green background, while the best AURAC is highlighted with an orange background. 
Gray shading indicates the model’s original accuracy on the corresponding benchmark.
}
\label{tab:understanding}
\resizebox{\textwidth}{!}{%
\begin{tabular}{ll c cc cc cc cc cc}
\toprule
& & & \multicolumn{2}{c}{Discrete Sem.\ Entropy} & \multicolumn{2}{c}{Semantic Entropy} & \multicolumn{2}{c}{P(True)} & \multicolumn{2}{c}{Predictive Entropy} & \multicolumn{2}{c}{Normalized Entropy} \\
\cmidrule(lr){4-5}\cmidrule(lr){6-7}\cmidrule(lr){8-9}\cmidrule(lr){10-11}\cmidrule(lr){12-13}
Benchmark & Model & \cellcolor{lightgray} Acc. & AUROC & AURAC & AUROC & AURAC & AUROC & AURAC & AUROC & AURAC & AUROC & AURAC \\
\midrule
\multirow{3}{*}{MMAU~\cite{sakshimmau}}
& Qwen-2.5-Omni-7B 
& \cellcolor{lightgray} 0.71 
& \cellcolor{lightgreen} 0.85 & \cellcolor{lightorange} 0.90 
& 0.84 & \cellcolor{lightorange} 0.90 
& 0.82 & 0.89 
& 0.64 & 0.81 
& 0.69 & 0.82 \\
& Qwen-2.5-Omni-3B 
& \cellcolor{lightgray} 0.73 
& 0.81 & 0.88 
& \cellcolor{lightgreen} 0.82 & \cellcolor{lightorange} 0.89 
& 0.78 & \cellcolor{lightorange} 0.89 
& 0.62 & 0.82 
& 0.66 & 0.84 \\
& Audio Flamingo 3              
& \cellcolor{lightgray} 0.74 
& 0.80 & 0.89 
& \cellcolor{lightgreen} 0.82 & \cellcolor{lightorange} 0.90 
& 0.72 & 0.86 
& 0.72 & 0.88 
& 0.73 & 0.88 \\
\midrule
\multirow{3}{*}{MMAR~\cite{ma2025mmar}}
& Qwen-2.5-Omni-7B 
& \cellcolor{lightgray} 0.58 
& 0.69 & 0.73 
& \cellcolor{lightgreen} 0.70 & 0.74 
& \cellcolor{lightgreen} 0.70 & \cellcolor{lightorange} 0.75 
& 0.59 & 0.68 
& 0.62 & 0.67 \\
& Qwen-2.5-Omni-3B 
& \cellcolor{lightgray} 0.58 
& 0.69 & 0.71 
& 0.70 & 0.71 
& \cellcolor{lightgreen} 0.71 & \cellcolor{lightorange} 0.74 
& 0.63 & 0.69 
& 0.62 & 0.68 \\
& Audio Flamingo 3              
& \cellcolor{lightgray} 0.55 
& 0.65 & \cellcolor{lightorange} 0.69 
& \cellcolor{lightgreen} 0.67 & \cellcolor{lightorange} 0.69 
& 0.61 & 0.65 
& 0.60 & 0.66 
& 0.58 & 0.64 \\
\midrule
\multirow{3}{*}{MMSU~\cite{wang2025mmsu}}
& Qwen-2.5-Omni-7B 
& \cellcolor{lightgray} 0.62 
& \cellcolor{lightgreen} 0.80 & \cellcolor{lightorange} 0.82 
& \cellcolor{lightgreen} 0.80 & 0.81 
& 0.78 & 0.81 
& 0.60 & 0.71 
& 0.69 & 0.76 \\
& Qwen-2.5-Omni-3B 
& \cellcolor{lightgray} 0.60 
& \cellcolor{lightgreen} 0.80 & \cellcolor{lightorange} 0.81 
& \cellcolor{lightgreen} 0.80 & \cellcolor{lightorange} 0.81 
& 0.79 & \cellcolor{lightorange} 0.81 
& 0.54 & 0.66 
& 0.68 & 0.75 \\
& Audio Flamingo 3              
& \cellcolor{lightgray} 0.55 
& 0.71 & 0.71 
& \cellcolor{lightgreen} 0.74 & \cellcolor{lightorange} 0.73 
& 0.69 & 0.71 
& 0.61 & 0.68 
& 0.64 & 0.68 \\
\midrule
\multirow{3}{*}{SAKURA~\cite{yang2025sakura}}
& Qwen-2.5-Omni-7B 
& \cellcolor{lightgray} 0.70 
& \cellcolor{lightgreen} 0.77 & \cellcolor{lightorange} 0.84 
& 0.76 & 0.83 
& 0.76 & 0.83 
& 0.63 & 0.78 
& 0.59 & 0.73 \\
& Qwen-2.5-Omni-3B 
& \cellcolor{lightgray} 0.69 
& 0.76 & 0.84 
& 0.77 & 0.84 
& \cellcolor{lightgreen} 0.79 & \cellcolor{lightorange} 0.87 
& 0.63 & 0.79 
& 0.60 & 0.74 \\
& Audio Flamingo 3              
& \cellcolor{lightgray} 0.62 
& 0.67 & 0.75 
& \cellcolor{lightgreen} 0.71 & \cellcolor{lightorange} 0.78 
& 0.69 & 0.75 
& 0.64 & 0.75 
& 0.58 & 0.71 \\
\bottomrule
\end{tabular}%
}
\end{table*}

\begin{table*}[t]
\centering
\caption{
Uncertainty estimation results on \textbf{Trustworthiness} benchmarks.
The best AUROC in each row is highlighted with a green background, while the best AURAC is highlighted with an orange background. 
Gray shading indicates the model’s original accuracy on the corresponding benchmark.
}
\label{tab:trustworthiness}
\resizebox{\textwidth}{!}{%
\begin{tabular}{ll c cc cc cc cc cc}
\toprule
& & & \multicolumn{2}{c}{Discrete Sem.\ Entropy} & \multicolumn{2}{c}{Semantic Entropy} & \multicolumn{2}{c}{P(True)} & \multicolumn{2}{c}{Predictive Entropy} & \multicolumn{2}{c}{Normalized Entropy} \\
\cmidrule(lr){4-5}\cmidrule(lr){6-7}\cmidrule(lr){8-9}\cmidrule(lr){10-11}\cmidrule(lr){12-13}
Benchmark & Model & \cellcolor{lightgray} Acc. & AUROC & AURAC & AUROC & AURAC & AUROC & AURAC & AUROC & AURAC & AUROC & AURAC \\
\midrule
\multirow{3}{*}{AQUA-Bench~\cite{kuan2026aqua}}
& Qwen-2.5-Omni-7B 
& \cellcolor{lightgray} 0.68 
& 0.70 & 0.80 
& 0.71 & 0.80 
& \cellcolor{lightgreen} 0.79 & \cellcolor{lightorange} 0.85 
& 0.55 & 0.73 
& 0.62 & 0.77 \\
& Qwen-2.5-Omni-3B 
& \cellcolor{lightgray} 0.76 
& 0.71 & 0.87 
& 0.72 & 0.87 
& 0.52 & 0.80 
& 0.72 & 0.88 
& \cellcolor{lightgreen} 0.75 & \cellcolor{lightorange} 0.89 \\
& Audio Flamingo 3              
& \cellcolor{lightgray} 0.18 
& 0.82 & 0.34 
& 0.83 & 0.34 
& \cellcolor{lightgreen} 0.89 & \cellcolor{lightorange} 0.40 
& 0.81 & 0.34 
& 0.81 & 0.34 \\
\midrule
\multirow{3}{*}{Hallucination~\cite{kuan2024understanding}}
& Qwen-2.5-Omni-7B 
& \cellcolor{lightgray} 0.89 
& \cellcolor{lightgreen} 0.78 & \cellcolor{lightorange} 0.95 
& \cellcolor{lightgreen} 0.78 & \cellcolor{lightorange} 0.95 
& 0.71 & \cellcolor{lightorange} 0.95 
& 0.66 & 0.93 
& 0.69 & 0.94 \\
& Qwen-2.5-Omni-3B 
& \cellcolor{lightgray} 0.89 
& 0.74 & \cellcolor{lightorange} 0.95 
& \cellcolor{lightgreen} 0.77 & \cellcolor{lightorange} 0.95 
& 0.55 & 0.93 
& 0.75 & \cellcolor{lightorange} 0.95 
& 0.71 & \cellcolor{lightorange} 0.95 \\
& Audio Flamingo 3              
& \cellcolor{lightgray} 0.80 
& 0.74 & 0.89 
& 0.75 & 0.89 
& 0.44 & 0.83 
& 0.75 & 0.91 
& \cellcolor{lightgreen} 0.78 & \cellcolor{lightorange} 0.92 \\
\bottomrule
\end{tabular}%
}
\end{table*}
\subsection{Overall Comparison of Uncertainty Estimation Methods}

Tables~\ref{tab:understanding} and~\ref{tab:trustworthiness} present the overall results of uncertainty estimation across benchmarks and models. 
Overall, we observe a clear pattern: semantic-based and verification-based methods, including discrete semantic entropy, semantic entropy, and P(True), generally outperform token-level baselines such as predictive entropy and normalized entropy. 
This trend is especially consistent on general audio understanding and reasoning benchmarks, where the strongest results are usually achieved by discrete semantic entropy, semantic entropy, or P(True).
Among the compared methods, semantic entropy and discrete semantic entropy are the most consistently strong performers. 
On MMAU, MMAR, MMSU, and SAKURA, these two methods frequently achieve the best or near-best AUROC across different models. 
P(True) is also competitive, and in some cases achieves the strongest results, such as on MMAR and SAKURA for Qwen2.5-Omni-3B and on AQUA for Qwen2.5-Omni-7B. 
These results suggest that uncertainty estimates operating at the semantic level, or through explicit self-verification, are generally more effective than methods based purely on token-level likelihood.

In contrast, predictive entropy is usually the weakest method, and normalized entropy, while sometimes improving over predictive entropy, still tends to underperform relative to semantic entropy, discrete semantic entropy, and P(True). 
This gap indicates that token-level uncertainty alone is often insufficient to capture answer correctness in audio-aware LLMs. 
In many cases, generated responses may be fluent and locally likely at the token level while still being semantically incorrect or weakly grounded in the input audio.
At the same time, the trustworthiness-oriented benchmarks reveal a more nuanced picture. 
While semantic-based and verification-based methods remain competitive overall, the relative ranking of methods becomes less stable across models and datasets. 
For example, on AQUA, P(True) performs best for Qwen2.5-Omni-7B, whereas normalized entropy performs best for Qwen2.5-Omni-3B, and P(True) shows particularly strong AUROC for Audio Flamingo 3. 
These results suggest that uncertainty estimation in trustworthiness-oriented settings is more model-dependent and benchmark-dependent.

\subsection{General Audio Understanding and Reasoning Benchmarks}

We next examine benchmark-specific results on general audio understanding and reasoning tasks, including MMAU, MMAR, MMSU, and SAKURA. 
As shown in Table~\ref{tab:understanding}, the overall trend is highly consistent across these benchmarks: semantic entropy, discrete semantic entropy, and P(True) generally outperform predictive entropy and normalized entropy. 
This suggests that uncertainty estimates based on semantic variation or explicit self-verification are more effective than token-level likelihood-based measures for identifying incorrect predictions in audio-aware LLMs.

Among these methods, semantic entropy and discrete semantic entropy are the most consistently strong performers. 
On MMAU, semantic entropy achieves the best AUROC for Qwen2.5-Omni-3B and Audio Flamingo 3, while discrete semantic entropy performs best for Qwen2.5-Omni-7B. 
On MMSU, semantic entropy and discrete semantic entropy again dominate across all three models. 
On SAKURA, P(True) becomes more competitive, achieving the best AUROC for Qwen2.5-Omni-3B, while semantic entropy and discrete semantic entropy remain strong for the other models. 
Across these benchmarks, predictive entropy is typically the weakest method, and normalized entropy, although sometimes stronger than predictive entropy, still generally falls behind the semantic- and verification-based approaches.
MMAR shows a slightly more mixed pattern, but still follows the same broad trend. 
For Qwen2.5-Omni-7B and Audio Flamingo 3, semantic entropy achieves the best AUROC, while P(True) performs best for Qwen2.5-Omni-3B. 
Even in this case, however, token-level entropy measures remain consistently weaker. 
This suggests that although the relative ranking among the strongest methods can vary by benchmark and model, semantic-based and verification-based uncertainty estimates are overall more robust than token-level baselines.

A more fine-grained breakdown of MMAR and MMSU into \textit{Perception} and \textit{Reasoning} subtasks is shown in Table~\ref{tab:subtask}. 
The subtask-level results reveal that the same high-level pattern largely holds within both categories, but also expose some interesting differences. 
On MMAR-Perception, P(True) is particularly competitive for the Qwen models, whereas on MMAR-Reasoning, discrete semantic entropy and semantic entropy become relatively stronger. 
On MMSU, discrete semantic entropy and semantic entropy remain highly competitive across both subtasks, while P(True) becomes especially strong on the reasoning subset for Qwen2.5-Omni-7B. 
These results suggest that uncertainty estimation behavior is not determined solely by the benchmark as a whole, but can also depend on whether the task is more perception-driven or reasoning-driven.
Overall, the results on general audio understanding and reasoning benchmarks show a consistent advantage for semantic-based and verification-based uncertainty estimates. 
At the same time, the subtask breakdown indicates that the relative strengths of these methods can shift depending on task structure.
% , motivating more detailed model-wise analysis in the next subsection.

\begin{table*}[t]
\centering
\caption{
Breakdown of MMAR and MMSU by \textbf{Perception} vs.\ \textbf{Reasoning} subtasks.
The best AUROC in each row is highlighted with a green background, while the best AURAC is highlighted with an orange background. 
Gray shading indicates the model’s original accuracy on the corresponding benchmark.
}
\label{tab:subtask}
\resizebox{\textwidth}{!}{%
\begin{tabular}{ll l c cc cc cc cc cc}
\toprule
& & & & \multicolumn{2}{c}{Discrete Sem.\ Entropy} & \multicolumn{2}{c}{Semantic Entropy} & \multicolumn{2}{c}{P(True)} & \multicolumn{2}{c}{Predictive Entropy} & \multicolumn{2}{c}{Normalized Entropy} \\
\cmidrule(lr){5-6}\cmidrule(lr){7-8}\cmidrule(lr){9-10}\cmidrule(lr){11-12}\cmidrule(lr){13-14}
Benchmark & Subtask & Model & \cellcolor{lightgray} Acc. & AUROC & AURAC & AUROC & AURAC & AUROC & AURAC & AUROC & AURAC & AUROC & AURAC \\
\midrule
\multirow{6}{*}{MMAR~\cite{ma2025mmar}}
& \multirow{3}{*}{Perception}
& Qwen-2.5-Omni-7B 
& \cellcolor{lightgray} 0.56 
& 0.68 & 0.70 
& 0.70 & 0.71 
& \cellcolor{lightgreen} 0.73 & \cellcolor{lightorange} 0.75 
& 0.61 & 0.67 
& 0.58 & 0.64 \\
& & Qwen-2.5-Omni-3B 
& \cellcolor{lightgray} 0.53 
& 0.68 & 0.67 
& 0.69 & 0.68 
& \cellcolor{lightgreen} 0.73 & \cellcolor{lightorange} 0.74 
& 0.62 & 0.65 
& 0.59 & 0.63 \\
& & Audio Flamingo 3              
& \cellcolor{lightgray} 0.51 
& 0.64 & 0.63 
& \cellcolor{lightgreen} 0.65 & \cellcolor{lightorange} 0.64 
& 0.63 & 0.63 
& 0.61 & 0.63 
& 0.56 & 0.58 \\
\cmidrule(lr){2-14}
& \multirow{3}{*}{Reasoning}
& Qwen-2.5-Omni-7B 
& \cellcolor{lightgray} 0.62 
& \cellcolor{lightgreen} 0.74 & \cellcolor{lightorange} 0.78 
& 0.73 & \cellcolor{lightorange} 0.78 
& 0.69 & 0.75 
& 0.59 & 0.71 
& 0.64 & 0.72 \\
& & Qwen-2.5-Omni-3B 
& \cellcolor{lightgray} 0.63 
& \cellcolor{lightgreen} 0.70 & \cellcolor{lightorange} 0.76 
& 0.69 & \cellcolor{lightorange} 0.76 
& 0.68 & \cellcolor{lightorange} 0.76 
& 0.62 & 0.72 
& 0.62 & 0.72 \\
& & Audio Flamingo 3              
& \cellcolor{lightgray} 0.62 
& 0.67 & 0.76 
& \cellcolor{lightgreen} 0.69 & \cellcolor{lightorange} 0.77 
& 0.58 & 0.70 
& 0.61 & 0.72 
& 0.61 & 0.72 \\
\midrule
\multirow{6}{*}{MMSU~\cite{wang2025mmsu}}
& \multirow{3}{*}{Perception}
& Qwen-2.5-Omni-7B 
& \cellcolor{lightgray} 0.45 
& \cellcolor{lightgreen} 0.73 & \cellcolor{lightorange} 0.66 
& 0.72 & 0.64 
& 0.72 & 0.64 
& 0.58 & 0.54 
& 0.58 & 0.54 \\
& & Qwen-2.5-Omni-3B 
& \cellcolor{lightgray} 0.45 
& \cellcolor{lightgreen} 0.71 & \cellcolor{lightorange} 0.63 
& 0.70 & \cellcolor{lightorange} 0.63 
& 0.69 & \cellcolor{lightorange} 0.63 
& 0.52 & 0.49 
& 0.58 & 0.54 \\
& & Audio Flamingo 3              
& \cellcolor{lightgray} 0.41 
& 0.64 & 0.55 
& \cellcolor{lightgreen} 0.66 & \cellcolor{lightorange} 0.56 
& \cellcolor{lightgreen} 0.66 & \cellcolor{lightorange} 0.56 
& 0.48 & 0.43 
& 0.50 & 0.46 \\
\cmidrule(lr){2-14}
& \multirow{3}{*}{Reasoning}
& Qwen-2.5-Omni-7B 
& \cellcolor{lightgray} 0.79 
& 0.78 & 0.91 
& 0.78 & 0.90 
& \cellcolor{lightgreen} 0.80 & \cellcolor{lightorange} 0.92 
& 0.58 & 0.84 
& 0.67 & 0.87 \\
& & Qwen-2.5-Omni-3B 
& \cellcolor{lightgray} 0.77 
& 0.81 & 0.91 
& \cellcolor{lightgreen} 0.82 & 0.91 
& 0.80 & \cellcolor{lightorange} 0.92 
& 0.56 & 0.81 
& 0.67 & 0.86 \\
& & Audio Flamingo 3              
& \cellcolor{lightgray} 0.69 
& 0.70 & 0.81 
& \cellcolor{lightgreen} 0.75 & \cellcolor{lightorange} 0.82 
& 0.69 & \cellcolor{lightorange} 0.82 
& 0.66 & \cellcolor{lightorange} 0.82 
& 0.62 & 0.79 \\
\bottomrule
\end{tabular}%
}
\end{table*}

\subsection{Trustworthiness-oriented Benchmarks}

We next examine the trustworthiness-oriented benchmarks, including AQUA-Bench~\cite{kuan2026aqua} and Audio-Hallucination~\cite{kuan2024understanding}. 
Compared with the general audio understanding and reasoning benchmarks, these tasks present a more challenging and nuanced setting for uncertainty estimation, as they more directly test whether the model can recognize unsupported, hallucinated, or unanswerable cases. 
As shown in Table~\ref{tab:trustworthiness}, the overall pattern becomes less uniform across models and benchmarks, indicating that uncertainty estimation in trustworthiness-oriented settings is more sensitive to both task structure and model behavior.

On AQUA-Bench, the relative ranking of uncertainty estimation methods varies substantially across models. 
For Qwen2.5-Omni-7B, P(True) achieves the best AUROC, suggesting that explicit self-verification is particularly effective for detecting unanswerable or unsupported predictions in this setting. 
In contrast, for Qwen2.5-Omni-3B, normalized entropy achieves the best AUROC, with predictive entropy also performing competitively. 
For Audio Flamingo 3, P(True) again performs strongly, while semantic entropy and predictive entropy also remain highly competitive. 
% This variability differs from the more stable trends observed on the general reasoning benchmarks, and suggests that uncertainty estimation for unanswerable audio questions may depend more strongly on the interaction between the benchmark and the model's answering behavior.
This variability contrasts with the more consistent trends on general reasoning benchmarks, suggesting that uncertainty estimation for unanswerable audio questions is more sensitive to each model’s behavior when the audio does not provide enough evidence.

The Audio-Hallucination benchmark shows a somewhat more stable pattern, but still exhibits clear model dependence. 
For the two Qwen models, semantic entropy and discrete semantic entropy achieve the strongest AUROC results, indicating that semantic-level uncertainty remains highly informative for detecting hallucinated audio-grounded answers. 
However, for Audio Flamingo 3, normalized entropy achieves the best AUROC, while predictive entropy is also competitive. 
This suggests that hallucination detection does not always favor the same uncertainty signal across different models, even when the benchmark itself is fixed.
Overall, the trustworthiness-oriented benchmarks reveal a more heterogeneous picture than the general understanding and reasoning benchmarks. 
Rather than showing a single consistently dominant method, they highlight stronger benchmark- and model-dependent variation in the usefulness of uncertainty estimates. 
This result is itself important: it suggests that uncertainty estimation for trustworthiness-oriented audio tasks may be inherently more difficult, and that conclusions drawn from general reasoning benchmarks do not always transfer directly to hallucination and unanswerable-question settings.

% \subsection{Model-wise Analysis}

\subsection{Application: Uncertainty-based Adaptive Inference}
\label{adaptive-inference}
Beyond evaluating uncertainty estimation as a diagnostic signal, we further explore whether it can be used to improve inference-time decision making. 
The key motivation is that not all examples require the same level of inference effort: some can be answered correctly with direct inference, while others are more challenging and may benefit from a more deliberate reasoning process~\cite{zhang2025adaptthink, li2025think}. 
Since reasoning-based inference is typically more expensive, it is undesirable to apply it uniformly to all inputs.
This makes uncertainty a natural routing signal. Let $u(x)$ denote the uncertainty score of the direct prediction for input $x$, and let $\tau$ denote a predefined threshold. 
We define adaptive inference as
\begin{equation}
\hat{y}_{\text{adaptive}}(x)=
\begin{cases}
\hat{y}_{\text{reason}}(x), & \text{if } u(x) > \tau, \\
\hat{y}_{\text{direct}}(x), & \text{otherwise},
\end{cases}
\end{equation}
where $\hat{y}_{\text{direct}}(x)$ is the answer produced by direct inference, and $\hat{y}_{\text{reason}}(x)$ is the answer produced by a more deliberate reasoning mode.
In our experiments, the reasoning mode follows a caption-then-reason strategy: the model is first prompted to describe the audio content, and is then asked to answer the question step by step based on that description. 
% This type of describe-then-reason decomposition has been explored in prior work~\cite{kuan2025can} on audio reasoning and timestamped audio captioning, where explicit intermediate descriptions are used to support downstream reasoning.
% The full prompting templates used in reasoning mode are presented in Table~\ref{tab:reasoning_inference_prompt}.
In our experiments, the reasoning mode follows a caption-then-reason strategy. 
The model is first prompted to describe the audio content, and then answers the question step by step. 
This describe-then-reason decomposition has been explored in prior work on audio reasoning and timestamped audio captioning, where explicit intermediate descriptions are used to support downstream reasoning~\cite{kuan2025can}. 
The full prompting templates used in reasoning mode are presented in Table~\ref{tab:reasoning_inference_prompt}.
We set the routing threshold to $\tau=0.25$. 
In preliminary experiments, we found that performance was relatively stable within a small range around this value, so we use $\tau=0.25$ as a fixed threshold in the main results.

Table~\ref{tab:adaptive} compares three inference strategies: direct inference, reasoning-based inference, and uncertainty-based adaptive inference. 
Overall, adaptive inference is beneficial in some settings, but its gains are not universal. 
In particular, it tends to help when the reasoning mode itself is competitive with or better than direct inference. 
For example, on MMAU and SAKURA, reasoning improves over direct inference for the Qwen models, and adaptive inference often matches or further improves upon the stronger strategy. 
On MMAU, adaptive inference achieves the best accuracy for both Qwen2.5-Omni-7B and Qwen2.5-Omni-3B, while on SAKURA it remains competitive with the best reasoning-based results.
We also report the relative token cost of adaptive inference compared to full reasoning.
Adaptive inference achieves comparable or better accuracy while using only 24–64\% of the tokens required by full reasoning.
Figure~\ref{fig:pareto-figure} presents a Pareto analysis comparing Reasoning and Adaptive inference, illustrating that Adaptive operating points consistently shift toward lower cost without proportional accuracy loss.
In contrast, when reasoning is not helpful, adaptive inference provides limited benefit and can even underperform direct inference. 
This is most clearly seen on MMAR, where reasoning substantially degrades performance for all three models, and adaptive inference correspondingly fails to improve over direct inference. 
A similar pattern is observed for Audio Flamingo 3 on MMAU, where direct inference is strongest and adaptive inference remains slightly below it. 
These results suggest that uncertainty alone is not sufficient to guarantee better routing decisions: adaptive inference can only be effective when the fallback reasoning mode is genuinely useful on uncertain examples.

This finding is also consistent with recent observations in the audio-aware LLM literature. 
Prior works~\cite{ma2025audio, xie2025audio, fan2025incentivizing, patel2025sound} have shown that simply introducing chain-of-thought or reasoning-style inference does not always improve performance in audio-aware LLMs, and can even lead to degraded results when the model is not properly trained for reasoning.
Our results reinforce this point from the perspective of uncertainty-aware routing: the usefulness of adaptive inference depends not only on the quality of the uncertainty estimate, but also on whether the alternative reasoning strategy is actually beneficial for the target benchmark and model.
% \begin{table}[t]
% \centering
% \caption{
% Accuracy under different inference strategies. 
% % Bold indicates the best result per row.
% Light blue shading indicates performance better than the Direct Answer baseline.
% }
% \label{tab:adaptive}
% \begin{tabular}{ll ccc}
% \toprule
% Benchmark & Model & Direct & Reasoning & Adaptive \\
% \midrule
% \multirow{3}{*}{MMAU~\cite{sakshimmau}}
% & Qwen-2.5-Omni-7B 
% & 0.71 & 0.75 & \cellcolor{lightblue} 0.76 \\
% & Qwen-2.5-Omni-3B 
% & 0.73 & 0.73 & \cellcolor{lightblue} 0.74 \\
% & Audio Flamingo 3              
% & 0.75 & 0.65 & 0.74 \\
% \midrule
% \multirow{3}{*}{MMAR~\cite{ma2025mmar}}
% & Qwen-2.5-Omni-7B 
% & 0.59 & 0.54 & 0.58 \\
% & Qwen-2.5-Omni-3B 
% & 0.58 & 0.53 & 0.57 \\
% & Audio Flamingo 3              
% & 0.56 & 0.47 & 0.53 \\
% \midrule
% \multirow{3}{*}{MMSU~\cite{wang2025mmsu}}
% & Qwen-2.5-Omni-7B 
% & 0.62 & 0.63 & 0.62 \\
% & Qwen-2.5-Omni-3B 
% & 0.60 & 0.61 & \cellcolor{lightblue} 0.62 \\
% & Audio Flamingo 3              
% & 0.56 & 0.54 & 0.56 \\
% \midrule
% \multirow{3}{*}{SAKURA~\cite{yang2025sakura}}
% & Qwen-2.5-Omni-7B 
% & 0.70 & 0.77 & \cellcolor{lightblue} 0.75 \\
% & Qwen-2.5-Omni-3B 
% & 0.69 & 0.73 & \cellcolor{lightblue} 0.73 \\
% & Audio Flamingo 3              
% & 0.63 & 0.70 & \cellcolor{lightblue} 0.70 \\
% \bottomrule
% \end{tabular}
% \end{table}

\begin{table}[t]
\centering
\caption{
Accuracy under different inference strategies. 
Light blue shading indicates performance better than the Direct Answer baseline. 
Values in parentheses denote the relative token cost of Adaptive inference compared to Reasoning.
}
\label{tab:adaptive}
\begin{tabular}{ll ccc}
\toprule
Benchmark & Model & Direct & Reasoning & Adaptive \\
\midrule
\multirow{3}{*}{MMAU}
& Qwen-2.5-Omni-7B 
& 0.71 & 0.75 & \cellcolor{lightblue} 0.76 (41\%) \\
& Qwen-2.5-Omni-3B 
& 0.73 & 0.73 & \cellcolor{lightblue} 0.74 (43\%) \\
& Audio Flamingo 3              
& 0.75 & 0.65 & 0.74 (24\%) \\
\midrule
\multirow{3}{*}{MMAR}
& Qwen-2.5-Omni-7B 
& 0.59 & 0.54 & 0.58 (61\%) \\
& Qwen-2.5-Omni-3B 
& 0.58 & 0.53 & 0.57 (64\%) \\
& Audio Flamingo 3              
& 0.56 & 0.47 & 0.53 (47\%) \\
\midrule
\multirow{3}{*}{MMSU}
& Qwen-2.5-Omni-7B 
& 0.62 & 0.63 & 0.62 (52\%) \\
& Qwen-2.5-Omni-3B 
& 0.60 & 0.61 & \cellcolor{lightblue} 0.62 (51\%) \\
& Audio Flamingo 3              
& 0.56 & 0.54 & 0.56 (54\%) \\
\midrule
\multirow{3}{*}{SAKURA}
& Qwen-2.5-Omni-7B 
& 0.70 & 0.77 & \cellcolor{lightblue} 0.75 (56\%) \\
& Qwen-2.5-Omni-3B 
& 0.69 & 0.73 & \cellcolor{lightblue} 0.73 (56\%) \\
& Audio Flamingo 3              
& 0.63 & 0.70 & \cellcolor{lightblue} 0.70 (61\%) \\
\bottomrule
\end{tabular}
\end{table}

\section{Discussion and Future Works}
Our study focuses on uncertainty estimation as an evaluation problem, but its value extends beyond benchmarking. 
More broadly, uncertainty estimates can serve as a useful interface between model predictions and downstream decision-making. 
Rather than treating every model output equally, a system equipped with reliable uncertainty estimates can decide when to trust its prediction, when to abstain, and when to invoke additional computation or verification. 
In this sense, uncertainty estimation is not only a diagnostic tool, but also a potential mechanism for making audio-aware LLMs more reliable in practice.
One immediate application is decision-making at inference time. 
As explored in our adaptive inference experiments (Section~\ref{adaptive-inference}), uncertainty can be used as a routing signal to decide whether an input should be handled by direct inference or a more expensive reasoning mode. 
More generally, uncertainty could be used as a threshold for abstention, fallback to external tools, or escalation to human review in high-stakes settings. 
This is especially relevant for audio-aware LLMs, where unsupported answers, hallucinated sound events, and failures on unanswerable questions can be difficult to detect from model outputs alone.

Another promising direction is to use uncertainty as an optimization signal during training. 
For example, uncertainty estimates could be used to prioritize difficult or ambiguous examples, reweight training instances, or guide data selection in self-training and reinforcement learning pipelines~\cite{chen2025seed}. 
In principle, such signals may provide a lightweight way to improve model reliability without requiring dense supervision. 
This is particularly attractive because uncertainty estimation is often unsupervised: it can be computed directly from model generations or self-verification signals, without requiring additional human annotations.
% At the same time, our results also suggest that uncertainty should not be viewed as a universally reliable solution. 
% Its usefulness depends on the task, the model, and the downstream application. 
% In particular, our adaptive inference results show that uncertainty-based routing is only effective when the alternative inference strategy is itself beneficial. 
At the same time, our results suggest that uncertainty should not be viewed as a universally reliable solution. 
Its value depends on the task, the model, and the downstream application. In particular, our adaptive inference results show that uncertainty-based routing improves performance only when the alternative inference strategy is itself stronger than the default strategy.
More broadly, uncertainty estimates in audio-aware LLMs may be affected by modality-specific factors such as perceptual ambiguity, weak audio grounding, and language priors. 
Future work should therefore investigate not only better uncertainty estimation methods, but also better ways of integrating uncertainty into training and inference.

\section{Conclusion}
This work presents the first systematic study of uncertainty estimation in audio-aware large language models. 
Our findings show that semantic-level and verification-based methods generally outperform token-level baselines, but this advantage is most stable on general reasoning tasks. 
On trustworthiness-oriented benchmarks, uncertainty estimation becomes more model- and task-dependent, suggesting that these settings pose fundamentally different challenges that deserve dedicated investigation. 
We hope this work provides a useful foundation for building more reliable audio-language systems.

\begin{figure}[t]
    \centering
    \includegraphics[width=0.5\textwidth]{figures/pareto-figure.png}
    \caption{
    Cost–accuracy Pareto frontier of Reasoning vs. Adaptive inference across four benchmarks.
    Each point represents a model under a fixed inference mode: hollow squares (Reasoning, 100\% token cost) and filled circles (Adaptive, reduced cost). 
    Dashed arrows indicate the shift from full reasoning to adaptive inference for each model.
    The gray dashed line represents the Pareto frontier, which consists of operating points that are not dominated in terms of both token cost and accuracy.
    }
    \label{fig:pareto-figure}
\end{figure}
\section{Limitations}

Our study has four main limitations. 
First, we primarily evaluate settings with relatively constrained answer spaces where correctness can be clearly defined. 
How well the studied uncertainty estimation methods generalize to open-ended audio-language tasks, such as free-form audio question answering, captioning, or dialogue, remains an open question for future work.
Second, the uncertainty estimation methods we study are largely inherited from the text LLM literature. 
While this design choice enables a controlled, method-agnostic comparison and helps establish baseline findings for the audio-language setting, these methods are not designed to explicitly model uncertainty arising from audio perception itself. 
Prior work in the vision-language domain~\cite{kostumov2024uncertainty, wang2026vlm} has begun to develop modality-aware uncertainty estimation methods tailored to visual perception; analogous efforts for audio-language models represent a promising research direction.
Third, we do not explore uncertainty signals derived from internal multimodal representations. 
For example, intermediate-layer attention patterns or hidden-state geometry may capture audio-specific ambiguity that token-level methods cannot surface. 
Investigating such representation-level signals could complement the output-level methods studied here and yield richer uncertainty estimates for audio-language models.
Finally, our adaptive inference experiments employ a threshold-based routing strategy with a fixed fallback reasoning mode. 
More sophisticated decision-making frameworks, such as learned routing policies, cascaded inference pipelines, or cost-aware selection mechanisms, could further improve the trade-off between computational cost and prediction reliability.

\bibliographystyle{IEEEtran}
\bibliography{refs}

\section{Generative AI Use Disclosure}
AI-assisted tools were used solely to improve the clarity and fluency of the manuscript.
All experiments, analyses, and related content were conducted and verified by the authors.

\section{Acknowledgments}
This work was supported by the Ministry of Education (MOE) of Taiwan under the project Taiwan Centers of Excellence in Artificial Intelligence, through the NTU Artificial Intelligence Center of Research Excellence (NTU AI-CoRE).

\section{Author Contributions}
Chun-Yi Kuan led the project, conducted the experiments, developed the analytical framework, and wrote the initial draft. 
Wei-Ping Huang conceived the original idea, provided technical guidance on experimental design, and contributed to drafting and revising the manuscript. 
Hung-yi Lee advised on research direction and experimental design, and provided critical feedback on the manuscript. 
All authors reviewed and approved the final version.
% All authors were involved throughout the research process, contributing to how the work was designed, conducted, and written up. 
% Chun-Yi Kuan led the project overall, running the experiments, developing the analytical framework, and writing most of the paper. 
% The initial idea was Wei-Ping Huang's, who also took part in drafting and refining the manuscript. 
% Wei-Ping Huang and Hung-yi Lee each brought deep technical knowledge to the table, playing a key role in shaping the research direction and providing critical feedback on both the experiments and the writing.

\section{Prompt Templates}

% \subsection{P(True) Verification Prompt}

For benchmark inference, we first use the basic inference prompt shown in Table~\ref{tab:basic_inference_prompt} to obtain the model's final prediction with low-temperature decoding. 
This prediction is used as the model's final answer for correctness evaluation. 
In the prompt, \texttt{\{question\}} is replaced by the input question, and the number of choices varies across questions.
For reasoning-mode inference in our uncertainty-based adaptive inference experiments, we use the prompt templates shown in Table~\ref{tab:reasoning_inference_prompt}.
For the self-verification P(True) baseline, we then ask the model to judge whether this generated answer is likely to be correct, given the input question and the generated answer. 
We use the prompt template shown in Table~\ref{tab:ptrue_prompt}, following previous work~\cite{kadavath2022language}.

% For the self-verification P(True) baseline, we use the prompt template shown in Table~\ref{tab:ptrue_prompt}, following previous work~\cite{kadavath2022language}.

\begin{table}[t]
\caption{Prompt used to obtain the model’s final prediction on the benchmark with low-temperature decoding. 
This prediction is used for correctness evaluation. \texttt{\{QUESTION\}} is replaced by the input question. 
The number of choices depends on each question.
}
\centering
\small
\begin{tabular}{p{0.95\linewidth}}
\toprule
\textbf{Basic Inference Prompt} \\
\midrule
\textbf{System Prompt:} \\
You are an audio assistant. Based on the given audio, select the best answer to the user's question from the provided choices. \\
\\
\textbf{User Prompt:} \\ 
Question: \texttt{\{QUESTION\}} \\
The answer could be: (a) \texttt{\{OPTION1\}} (b) \texttt{\{OPTION2\}} (c) \texttt{\{OPTION3\}} (d) \texttt{\{OPTION4\}} \\
\bottomrule
\end{tabular}
\label{tab:basic_inference_prompt}
\end{table}
\begin{table}[htbp]
\caption{Prompt used for the reasoning mode in the uncertainty-based adaptive inference experiments. 
\texttt{{QUESTION}} is replaced with the input question, and the number of choices varies depending on the question.
}
\centering
\small
\begin{tabular}{p{0.95\linewidth}}
\toprule
\textbf{Inference Prompt for Reasoning Mode} \\
\midrule
\textbf{System Prompt:} \\
You are an audio reasoning assistant. \\
\\
First, describe the audio based only on audible evidence. 
Summarize the main sound events, speakers, acoustic scene, and any relevant temporal order. 
Do not use outside knowledge or guess beyond what can be heard. \\
\\
Then, answer the multiple-choice question using only the audio evidence. 
Choose exactly one option from the provided choices. \\ \\
Output in the following format: \\
Audio Caption: \texttt{<your caption>} \\
Reasoning: \texttt{<brief reasoning based on the audio>} \\
Answer: \texttt{<one option only>} \\
\\
\textbf{User Prompt:} \\ 
Question: \texttt{\{QUESTION\}} \\
The answer could be: (a) \texttt{\{OPTION1\}} (b) \texttt{\{OPTION2\}} (c) \texttt{\{OPTION3\}} (d) \texttt{\{OPTION4\}} \\
\bottomrule
\end{tabular}
\label{tab:reasoning_inference_prompt}
\end{table}
\begin{table}[hbpt]
\vspace{4pt}
\caption{Verification prompt used for self-verification in P(True), where \texttt{\{question\}} and \texttt{\{answer\}} are replaced with the corresponding question and answer.}
\centering
\small
\begin{tabular}{p{0.95\linewidth}}
\toprule
\textbf{Verification Prompt} \\
\midrule
\textbf{System Prompt:} \\
You are a careful audio verification assistant. Listen to the provided audio and judge whether the answer to the question is directly supported by audible evidence in the audio. Do not rely on plausibility, common sense, or outside knowledge. Respond with only one word: true or false. \\
\\
\textbf{User Prompt:} \\ 
Question: \texttt{\{QUESTION\}} \\
Proposed Answer: \texttt{\{ANSWER\}} \\
Is the proposed answer true or false based on the audio? \\
Answer with only one word: true or false. \\
Answer: \\
\bottomrule
\end{tabular}
% \vspace{4pt}
% \caption{Verification prompt used for self-verification in P(True), where \texttt{\{question\}} and \texttt{\{answer\}} are replaced with the corresponding question and answer.}
\label{tab:ptrue_prompt}
\end{table}
\begin{table}[t]
% \vspace{4pt}
\caption{Self-assessment prompt used in P(True) to estimate whether the model can answer the input question correctly, where \texttt{\{question\}} is replaced with the input question.}
\centering
\small
\begin{tabular}{p{0.95\linewidth}}
\toprule
\textbf{Self-Assessment Prompt} \\
\midrule
\textbf{System Prompt:} \\
You are an assistant that evaluates its own ability to answer questions. Given a question, decide whether you are confident that you can answer the question correctly. \\
\\
\textbf{User Prompt:} \\ 
Question: \texttt{\{QUESTION\}} \\
Are you able to answer the question correctly? \\
Answer with only a single word: yes or no. \\
Answer: \\
\bottomrule
\end{tabular}
\label{tab:self_assessment_prompt}
\end{table}

% \subsubsection{Formulation}
% Following prior work~\cite{yang2026calibration} on uncertainty estimation, we define capability calibration as the alignment between a model’s predicted confidence and its \textit{expected correctness} on a given input.
% Unlike response calibration, where correctness is binary (0 or 1) for a single prediction, capability calibration considers the model’s expected accuracy over its output distribution.

% For each input $x$, we estimate the expected correctness as:
% \[
% \mathbb{E}_{y \sim p(y|x)}[\mathbb{1}(\text{correct}(y))]
% \]
% which corresponds to the probability that the model would answer correctly when sampling from its predictive distribution.

% \paragraph{Estimation.}
% In practice, we approximate this expectation via Monte Carlo sampling.
% For each input, we draw $K=100$ stochastic samples from $p(y|x)$ using high-temperature decoding, and compute the fraction of correct responses.
% This yields a soft target in $[0,1]$, which we treat as the ground-truth expected accuracy.

% To estimate the model’s predicted confidence, we use the P(True) method.
% Specifically, given an input $x$, we prompt the model to predict whether it would answer the question correctly, and compute the confidence score from the normalized likelihood of the binary outputs (true/false).
% The prompt template is shown in Table~X.

% \paragraph{Evaluation.}
% We evaluate capability calibration by measuring how well the predicted confidence aligns with the estimated expected accuracy.

\section{Capability Calibration}

In this work, we primarily focus on \textit{response calibration}, which evaluates whether a model’s confidence aligns with the correctness of its generated answer.
A related but distinct problem is \textit{capability calibration}~\cite{yang2026calibration, kadavath2022language}: whether a model can anticipate, \emph{prior to generating an answer}, whether it will respond correctly.
This setting is particularly relevant for decision-making scenarios such as selective prediction and abstention, where the model must decide whether to attempt an answer at all.
Since capability calibration requires a different evaluation protocol and is not the main focus of this work, we do not include it in the main experiments.
Instead, we provide a preliminary analysis of the Qwen2.5-Omni 7B and 3B variants in the following subsections.

\subsection{Formulation}

Following prior work on capability calibration~\cite{yang2026calibration}, we define capability calibration as the alignment between a model’s predicted confidence and its expected correctness under the predictive distribution.
Formally, given an input $x$ consisting of an audio clip and a corresponding instruction or question, let $p(y \mid x)$ denote the model's predictive distribution over outputs. The expected correctness is defined as:
\[
\mathbb{E}_{y \sim p(y \mid x)} \left[ \mathbf{1}(\text{correct}(y)) \right],
\]
which represents the probability that a randomly sampled response from the model is correct.
Let $c(x) \in [0,1]$ denote the model’s predicted confidence of answering correctly. Capability calibration then evaluates how well $c(x)$ aligns with the expected correctness:
\[
c(x) \approx \mathbb{E}_{y \sim p(y \mid x)} \left[ \mathbf{1}(\text{correct}(y)) \right].
\]
% The prompt templates designed for confidence estimation is shown in Table~\ref{tab:self_assessment_prompt}.

\subsection{Estimation}
In practice, the expected correctness is intractable to compute exactly, since the full predictive distribution \(p(y \mid x)\) is not directly accessible. 
We therefore approximate it using Monte Carlo sampling. For each input \(x\), we draw \(K=100\) stochastic samples \(\{y_k\}_{k=1}^K\) from \(p(y \mid x)\) using high-temperature decoding, and compute
\[
\hat{a}(x) = \frac{1}{K} \sum_{k=1}^{K} \mathbf{1}(\text{correct}(y_k)).
\]
Here, \(\hat{a}(x)\) is the empirical accuracy over sampled responses, which serves as a Monte Carlo estimate of the expected correctness:
\[
\mathbb{E}_{y \sim p(y \mid x)} \left[ \mathbf{1}(\text{correct}(y)) \right].
\]
As such, \(\hat{a}(x)\) provides a soft target in \([0,1]\) for evaluating capability calibration.
To estimate the model’s predicted confidence \(c(x)\), we adopt a method similar to P(True). 
Specifically, we prompt the model to assess whether it would answer the question correctly, and compute the confidence score from the normalized likelihood of binary outputs (true/false). 
In previous work~\cite{kadavath2022language}, this is referred to as P(IK), representing the probability that the model assigns to ``I know (IK)''.
The prompt template used to estimate the model’s confidence in answering the question correctly is shown in Table~\ref{tab:self_assessment_prompt}.

\subsection{Evaluation}
% We evaluate capability calibration by measuring the alignment between the predicted confidence $c(x)$ and the estimated expected correctness $\hat{a}(x)$. 
% This can be quantified using standard calibration metrics such as Expected Calibration Error (ECE) and Brier score. 
We evaluate capability calibration by measuring the alignment between the predicted confidence $c(x)$ and the estimated expected correctness $\hat{a}(x)$ using Expected Calibration Error (ECE) and Brier score.
For ECE, we adopt equal-mass binning. Specifically, we first sort all inputs by predicted confidence and partition them into $M$ bins with approximately equal numbers of samples. Let $B_m$ denote the set of inputs in the $m$-th bin. ECE is then computed as
\[
\mathrm{ECE} = \sum_{m=1}^{M} \frac{|B_m|}{N}
\left|
\frac{1}{|B_m|}\sum_{x \in B_m} c(x)
-
\frac{1}{|B_m|}\sum_{x \in B_m} \hat{a}(x)
\right|.
\]

We also report the Brier score, defined as
\[
\mathrm{Brier} = \frac{1}{N}\sum_{x}\bigl(c(x)-\hat{a}(x)\bigr)^2.
\]

To visualize calibration quality, we plot equal-mass reliability diagrams, where predictions are sorted by confidence and partitioned into bins of approximately equal sample size. 
We adopt equal-mass binning rather than equal-width binning because the latter can produce bins with very few or no samples in regions of low density, leading to noisy and unreliable estimates of per-bin accuracy. 
Each bin is represented as a point whose $x$-coordinate is the mean predicted confidence and $y$-coordinate is the mean empirical accuracy. 
Points closer to the diagonal ($y = x$) indicate better calibration. 
We report ECE and Brier score alongside each diagram.

\subsection{Results}
% \begin{table}[t]
% \centering
% \small
% \caption{
% Capability calibration results of Qwen2.5-Omni-7B across different benchmarks. 
% Lower is better for both Expected Calibration Error (ECE) and Brier score.
% }
% \label{tab:capability_calibration_qwen7b}
% \begin{tabular}{llcc}
% \toprule
% Benchmark & Category & ECE $\downarrow$ & Brier $\downarrow$ \\
% \midrule
% \multirow{4}{*}{MMAU}
% & Sound   & 0.042 & 0.108 \\
% & Speech  & 0.070 & 0.115 \\
% & Music   & 0.090 & 0.115 \\
% & Overall & 0.054 & 0.112 \\
% \midrule
% \multirow{3}{*}{MMAR}
% & Perception & 0.116 & 0.145 \\
% & Reasoning  & 0.121 & 0.107 \\
% & Overall    & 0.114 & 0.124 \\
% \midrule
% \multirow{3}{*}{MMSU}
% & Perception & 0.212 & 0.129 \\
% & Reasoning  & 0.041 & 0.103 \\
% & Overall    & 0.108 & 0.117 \\
% \midrule
% \multirow{5}{*}{SAKURA}
% & Animal   & 0.119 & 0.070 \\
% & Emotion  & 0.227 & 0.170 \\
% & Gender   & 0.169 & 0.125 \\
% & Language & 0.141 & 0.073 \\
% & Overall  & 0.077 & 0.110 \\
% \bottomrule
% \end{tabular}
% \end{table}

\begin{table}[t]
\centering
\small
\caption{
Capability calibration results of Qwen2.5-Omni models (7B and 3B variants) across benchmarks.
Lower is better for both Expected Calibration Error (ECE) and Brier score (Brier).
The best ECE in each row is highlighted with a green background, while the best Brier score is highlighted with an orange background. 
}
\label{tab:capability_calibration_qwen7b}
\begin{tabular}{llcccc}
\toprule
& & \multicolumn{2}{c}{7B} & \multicolumn{2}{c}{3B} \\
\cmidrule(lr){3-4} \cmidrule(lr){5-6}
Benchmark & Category & ECE $\downarrow$ & Brier $\downarrow$ & ECE $\downarrow$ & Brier $\downarrow$ \\
\midrule
\multirow{4}{*}{MMAU}
& Sound   
& \cellcolor{lightgreen} 0.042 & \cellcolor{lightorange} 0.108 
& 0.068 & 0.112 \\
& Speech  
& 0.070 & 0.115 
& \cellcolor{lightgreen} 0.067 & \cellcolor{lightorange} 0.101 \\
& Music   
& \cellcolor{lightgreen} 0.090 & \cellcolor{lightorange} 0.115 
& 0.091 & 0.122 \\
& Overall 
& \cellcolor{lightgreen} 0.054 & \cellcolor{lightorange} 0.112 
& 0.104 & 0.114 \\
\midrule
\multirow{3}{*}{MMAR}
& Perception 
& \cellcolor{lightgreen} 0.116 & 0.145 
& 0.141 & \cellcolor{lightorange} 0.128 \\
& Reasoning  
& \cellcolor{lightgreen} 0.121 & \cellcolor{lightorange} 0.107 
& 0.131 & 0.138 \\
& Overall    
& \cellcolor{lightgreen} 0.114 & 0.124 
& 0.164 & \cellcolor{lightorange} 0.118 \\
\midrule
\multirow{3}{*}{MMSU}
& Perception 
& 0.212 & 0.129 
& \cellcolor{lightgreen} 0.093 & 0.109 \\
& Reasoning  
& \cellcolor{lightgreen} 0.041 & \cellcolor{lightorange} 0.103 
& 0.170 & 0.108 \\
& Overall    
& 0.108 & 0.117 
& \cellcolor{lightgreen} 0.044 & \cellcolor{lightorange} 0.110 \\
\midrule
\multirow{5}{*}{SAKURA}
& Animal   
& \cellcolor{lightgreen} 0.119 & \cellcolor{lightorange} 0.070 
& 0.131 & 0.110 \\
& Emotion  
& 0.227 & 0.170 
& \cellcolor{lightgreen} 0.181 & \cellcolor{lightorange} 0.167 \\
& Gender   
& 0.169 & 0.125 
& \cellcolor{lightgreen} 0.138 & \cellcolor{lightorange} 0.122 \\
& Language 
& 0.141 & \cellcolor{lightorange} 0.073 
& \cellcolor{lightgreen} 0.134 & 0.093 \\
& Overall  
& \cellcolor{lightgreen} 0.077 & \cellcolor{lightorange} 0.110 
& 0.092 & 0.123 \\
\bottomrule
\end{tabular}
\end{table}

Table~\ref{tab:capability_calibration_qwen7b} reports the Expected Calibration Error (ECE) and Brier score for the Qwen2.5-Omni 7B and 3B models across four benchmarks. 
Overall, neither model consistently dominates the other in calibration quality. 
On MMAU, the 7B model achieves lower ECE overall (0.054 vs. 0.104), while the 3B model yields comparable or slightly better Brier scores in certain categories such as Speech (0.101 vs. 0.115). 
On MMAR, the 7B model tends to have lower ECE overall (0.114 vs. 0.164), though the 3B model obtains a lower Brier score on the Overall split (0.118 vs. 0.124). 
A notable case is MMSU, where the 3B model outperforms the 7B model in overall ECE (0.044 vs. 0.108), yet the 7B model achieves a lower Brier score on Reasoning (0.103 vs. 0.108). 
On SAKURA, the two models show relatively similar calibration, with the 7B model achieving a slightly lower overall ECE (0.077 vs. 0.092) and Brier score (0.110 vs. 0.123). 
These results suggest that model size does not uniformly improve calibration, and the relative performance varies across benchmarks and task categories.

\begin{figure*}[t]
    \centering
    \includegraphics[width=\textwidth]{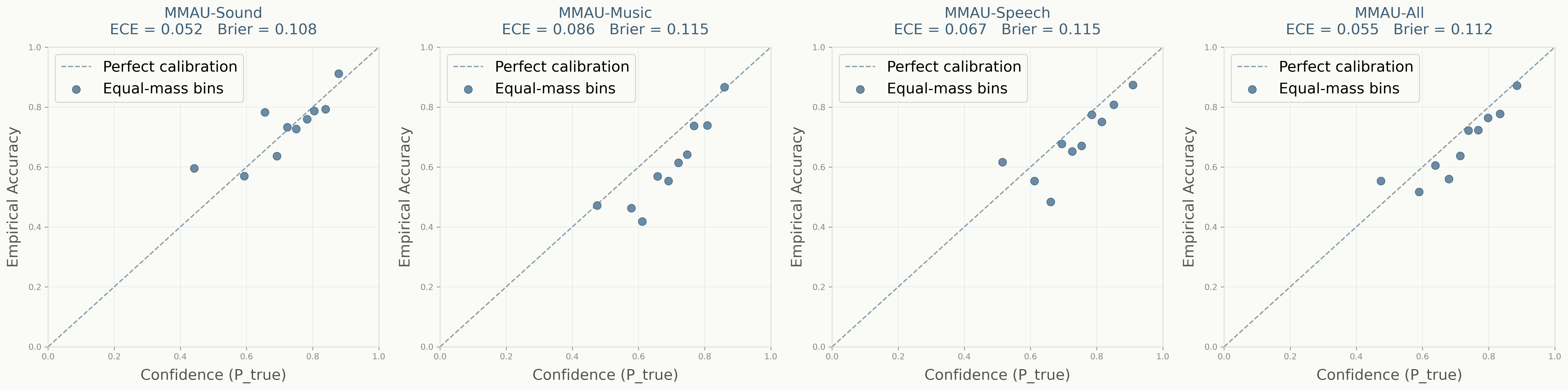}
    \caption{
    Reliability diagrams for \textbf{Qwen2.5-Omni-7B}~\cite{xu2025qwen2} on MMAU~\cite{sakshimmau}, reported for both the overall benchmark and three domains: sound, speech, and music.
    % The dashed line indicates perfect calibration. 
    Each point corresponds to an equal-mass bin. 
    Lower ECE and Brier score indicate better calibration.
    }
    \label{fig:mmau_calibration}
\end{figure*}

\begin{figure*}[t]
    \centering
    \includegraphics[width=\textwidth]{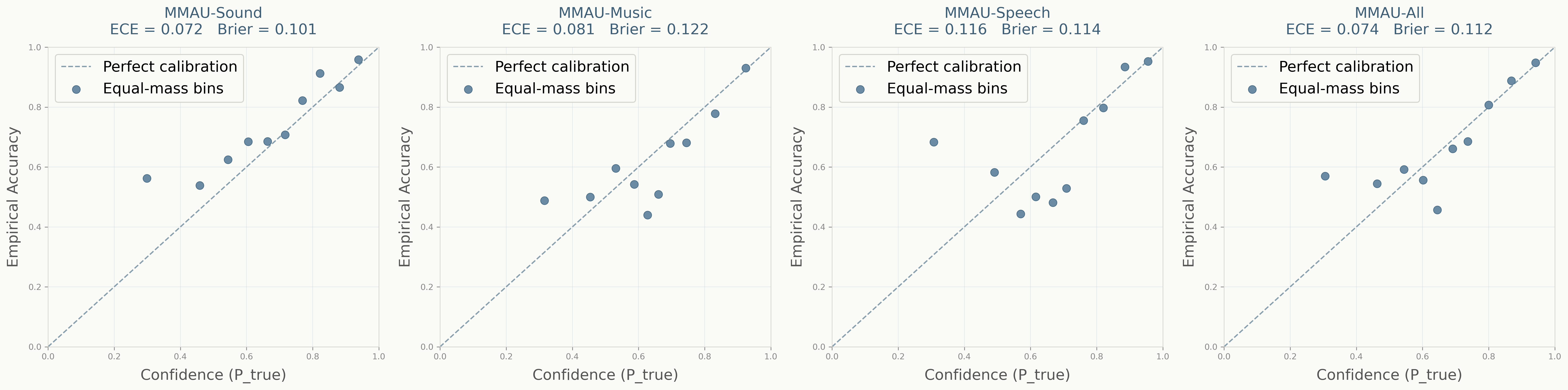}
    \caption{
    Reliability diagrams for \textbf{Qwen2.5-Omni-3B}~\cite{xu2025qwen2} on MMAU~\cite{sakshimmau}, reported for both the overall benchmark and three domains: sound, speech, and music.
    % The dashed line indicates perfect calibration. 
    Each point corresponds to an equal-mass bin. 
    Lower ECE and Brier score indicate better calibration.
    }
    \label{fig:mmau_calibration_3b}
\end{figure*}

\begin{figure*}[t]
    \centering
    \includegraphics[width=\textwidth]{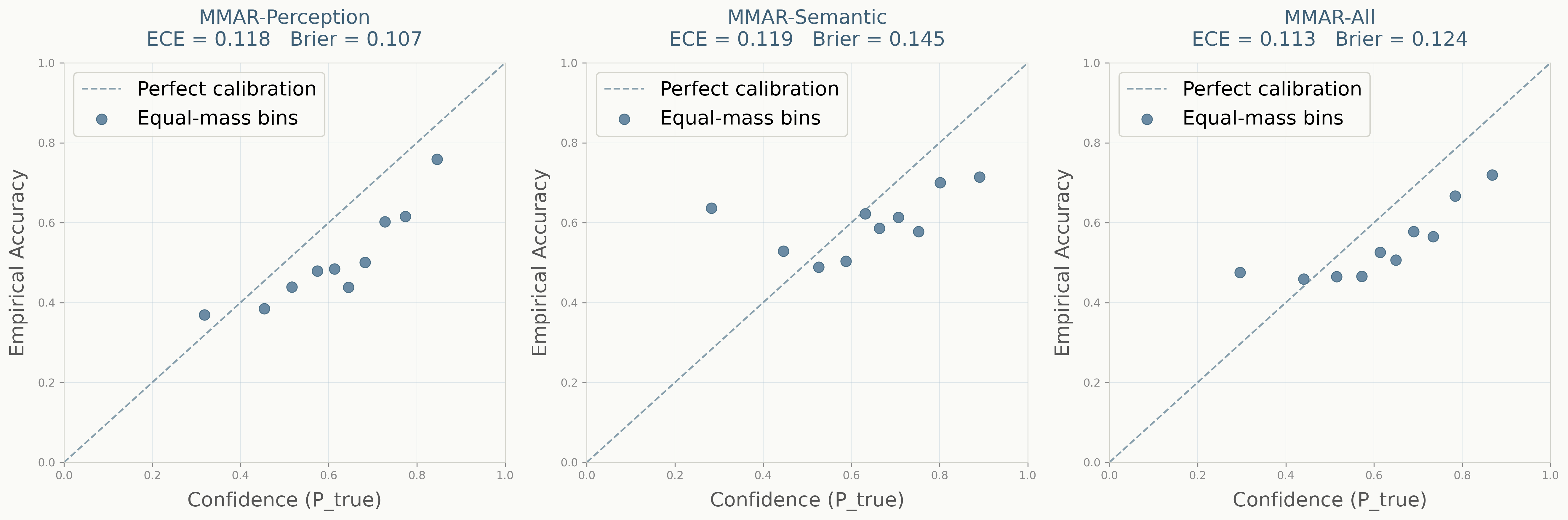}
    \caption{
    Reliability diagrams for \textbf{Qwen2.5-Omni-7B}~\cite{xu2025qwen2} on MMAR~\cite{ma2025mmar}, reported for both the overall benchmark and the two benchmark-defined task categories: Perception and Semantic.
    % The dashed line indicates perfect calibration. 
    Each point corresponds to an equal-mass bin. 
    Lower ECE and Brier score indicate better calibration.
    }
    \label{fig:mmar_calibration}
\end{figure*}

\begin{figure*}[t]
    \centering
    \includegraphics[width=\textwidth]{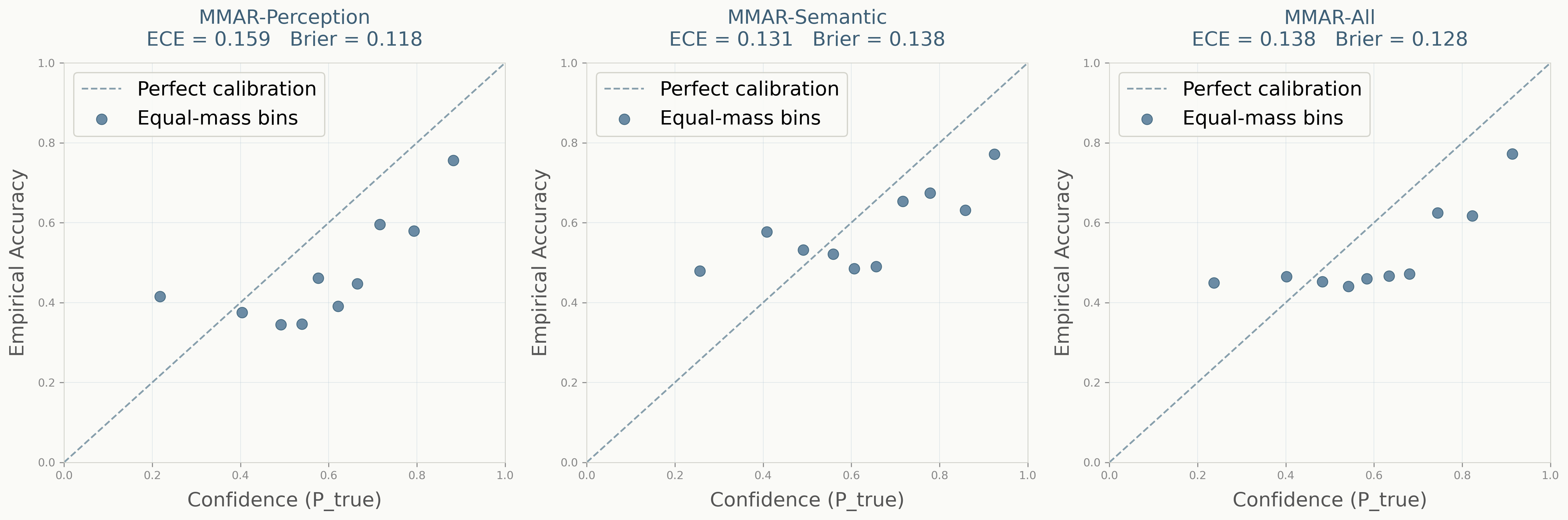}
    \caption{
    Reliability diagrams for \textbf{Qwen2.5-Omni-3B}~\cite{xu2025qwen2} on MMAR~\cite{ma2025mmar}, reported for both the overall benchmark and the two benchmark-defined task categories: Perception and Semantic.
    % The dashed line indicates perfect calibration. 
    Each point corresponds to an equal-mass bin. 
    Lower ECE and Brier score indicate better calibration.
    }
    \label{fig:mmar_calibration_3b}
\end{figure*}

\begin{figure*}[t]
    \centering
    \includegraphics[width=\textwidth]{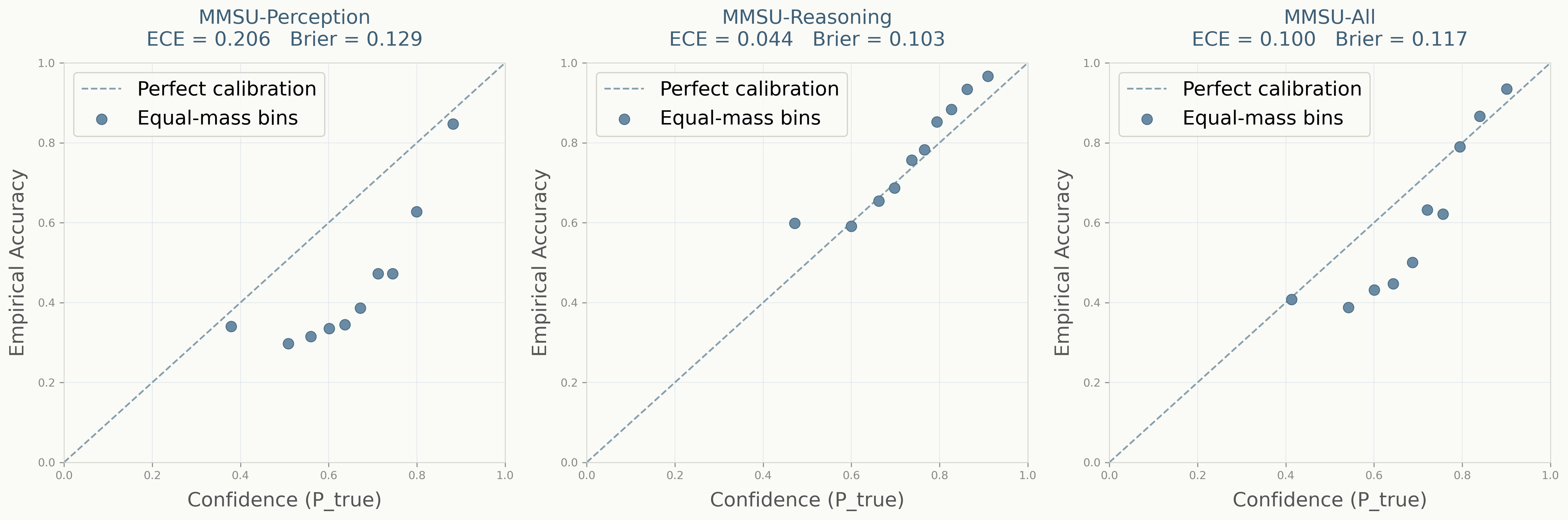}
    \caption{
    Reliability diagrams for \textbf{Qwen2.5-Omni-7B}~\cite{xu2025qwen2} on MMSU~\cite{wang2025mmsu}, reported for both the overall benchmark and the two task categories: Perception and Reasoning.
    % The dashed line indicates perfect calibration. 
    Each point corresponds to an equal-mass bin. 
    Lower ECE and Brier score indicate better calibration.
    }
    \label{fig:mmsu_calibration}
\end{figure*}

\begin{figure*}[hbpt]
    \centering
    \includegraphics[width=\textwidth]{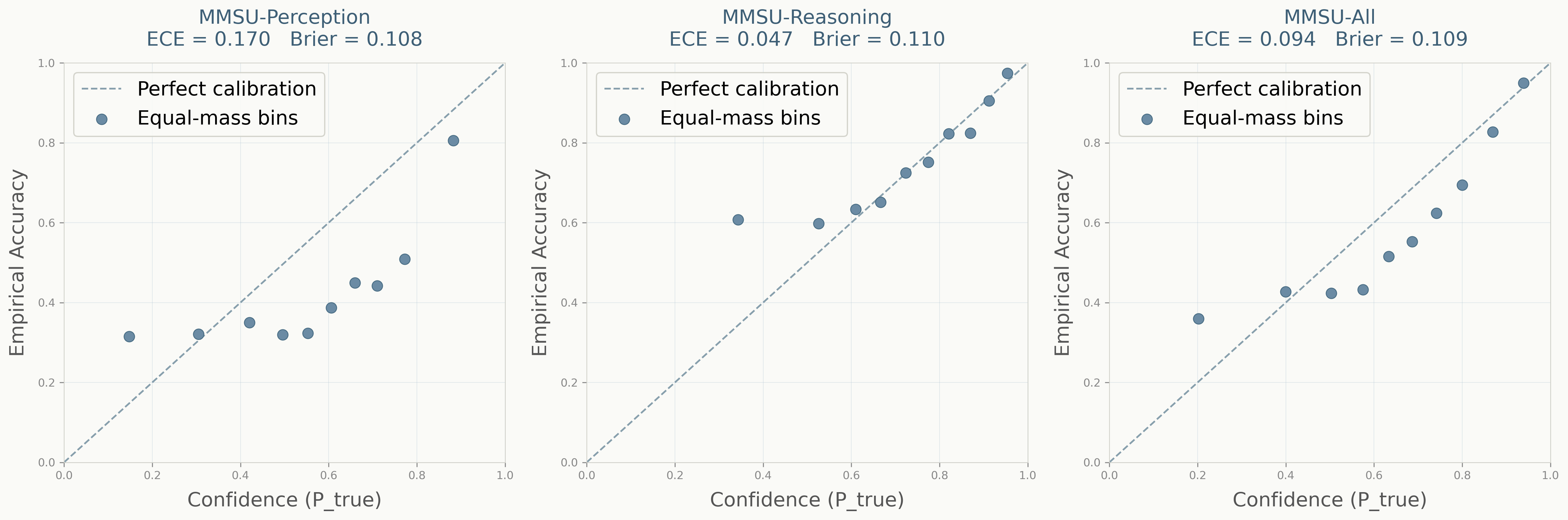}
    \caption{
    Reliability diagrams for \textbf{Qwen2.5-Omni-3B}~\cite{xu2025qwen2} on MMSU~\cite{wang2025mmsu}, reported for both the overall benchmark and the two task categories: Perception and Reasoning.
    % The dashed line indicates perfect calibration. 
    Each point corresponds to an equal-mass bin. 
    Lower ECE and Brier score indicate better calibration.
    }
    \label{fig:mmsu_calibration_3b}
\end{figure*}

\begin{figure*}[t]
    \centering
    \includegraphics[width=\textwidth]{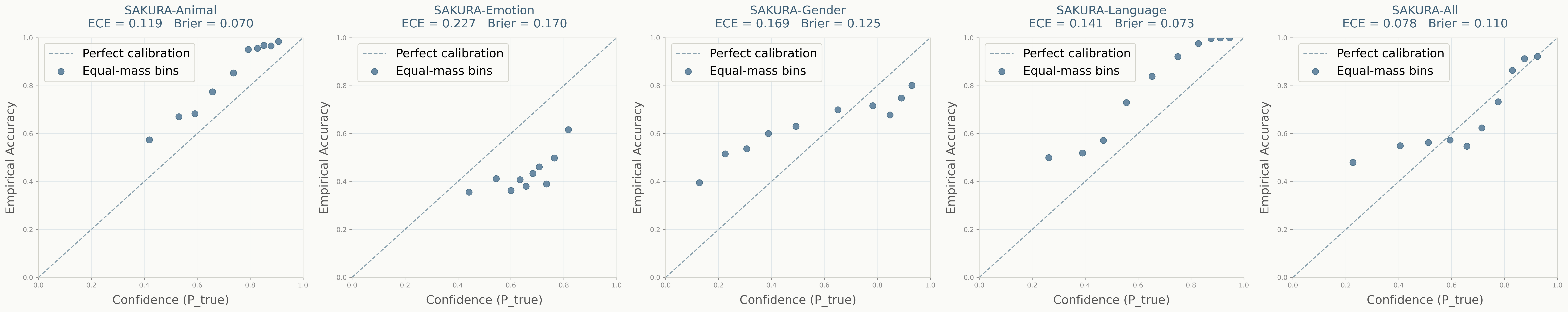}
    \caption{
    Reliability diagrams for \textbf{Qwen2.5-Omni-7B}~\cite{xu2025qwen2} on SAKURA~\cite{yang2025sakura}, including overall performance and four tracks: animal, emotion, language, and gender.
    % The dashed line indicates perfect calibration. 
    Each point corresponds to an equal-mass bin. 
    Lower ECE and Brier score indicate better calibration.
    }
    \label{fig:sakura_calibration}
\end{figure*}

\begin{figure*}[t]
    \centering
    \includegraphics[width=\textwidth]{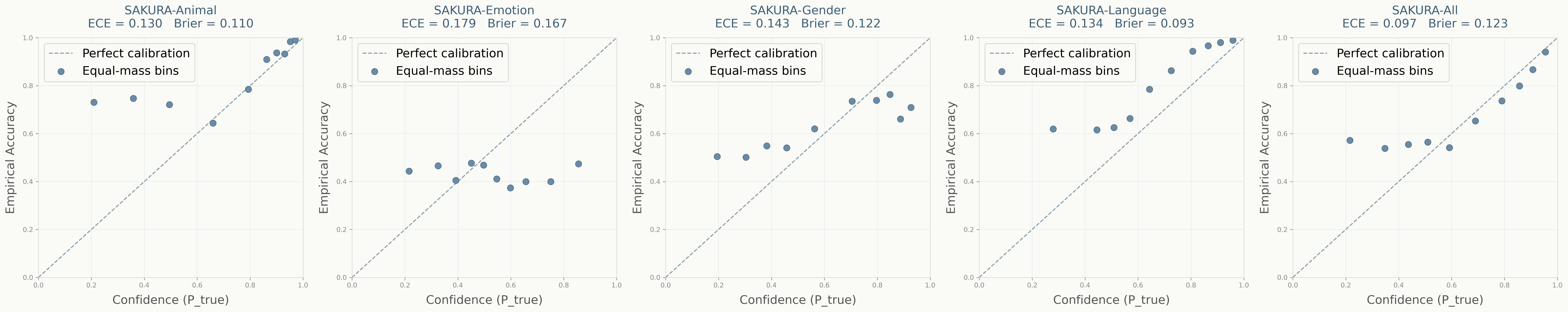}
    \caption{
    Reliability diagrams for \textbf{Qwen2.5-Omni-3B}~\cite{xu2025qwen2} on SAKURA~\cite{yang2025sakura}, including overall performance and four tracks: animal, emotion, language, and gender.
    % The dashed line indicates perfect calibration. 
    Each point corresponds to an equal-mass bin. 
    Lower ECE and Brier score indicate better calibration.
    }
    \label{fig:sakura_calibration_3b}
\end{figure*}

The reliability diagrams~\Cref{fig:mmau_calibration,fig:mmau_calibration_3b,fig:mmar_calibration,fig:mmar_calibration_3b,fig:mmsu_calibration,fig:mmsu_calibration_3b,fig:sakura_calibration,fig:sakura_calibration_3b} and calibration metrics (Table~\ref{tab:capability_calibration_qwen7b}) for Qwen2.5-Omni-7B and 3B models reveal that capability calibration quality varies across benchmarks and task categories. 
For brevity, we focus our discussion on Qwen2.5-Omni-7B, noting that the 3B model shows broadly similar patterns.
Among the four benchmarks, MMAU exhibits the best overall calibration (ECE = 0.054, Brier = 0.112), with all three domains—sound, speech, and music—showing relatively tight alignment between predicted confidence and empirical accuracy, as reflected in the near-diagonal bins in Figure~\ref{fig:mmau_calibration}. 
In contrast, MMAR and MMSU show higher calibration error overall (Table~\ref{tab:capability_calibration_qwen7b}), suggesting that the model's self-assessed ability to answer correctly becomes less reliable as task difficulty increases. 
A particularly notable pattern emerges when comparing perception and reasoning subtasks. 
As shown in Table~\ref{tab:capability_calibration_qwen7b} and Figure~\ref{fig:mmsu_calibration}, the perception subset of MMSU exhibits markedly poor calibration (ECE = 0.212), whereas the reasoning subset is well-calibrated (ECE = 0.041). 
This asymmetry suggests that the model systematically overestimates its perceptual capabilities while maintaining more accurate self-assessment on reasoning-oriented questions. 
On MMAR (Figure~\ref{fig:mmar_calibration}), this gap is less pronounced, with both subtasks showing comparable ECE values around 0.12. 
Within SAKURA (Figure~\ref{fig:sakura_calibration} and Table~\ref{tab:capability_calibration_qwen7b}), domain-level variation is also evident: the emotion track yields the highest calibration error (ECE = 0.227), while animal and language tracks are better calibrated. 
This is consistent with the intuition that subjective or perceptually ambiguous audio attributes pose greater challenges not only for prediction accuracy but also for the model's ability to anticipate its own correctness. 
Across most benchmarks and domains, the reliability diagrams indicate a tendency toward overconfidence, with many equal-mass bins falling below the diagonal in the high-confidence region. 
These findings complement our main results on response calibration by showing that the task- and modality-dependent nature of uncertainty in audio-aware LLMs extends to the capability calibration setting as well.




\end{document}